\documentclass[aps,prd,amssymb,amsmath,nofootinbib]{revtex4}

\begin{document}

\title{Metric fluctuations of an evaporating black hole
from back reaction of stress tensor fluctuations}
\author{B.~L. Hu}
\affiliation{Maryland Center for Fundamental Physics, Department of
Physics, University of Maryland, College Park, Maryland 20742-4111}
\author{Albert Roura}
\affiliation{Theoretical Division, T-8, Los Alamos National Laboratory,
M.S.~B285, Los Alamos, NM 87545}

\begin{abstract}

This paper delineates the first steps in a systematic quantitative
study of the spacetime fluctuations induced by quantum fields in an
evaporating black hole under the stochastic gravity program. The
central object of interest is the noise kernel, which is the
symmetrized two-point quantum correlation function of the stress
tensor operator.
As a concrete example we apply it to the study of the
spherically-symmetric sector of metric perturbations around an
evaporating black hole background geometry. For macroscopic black
holes we find that those fluctuations grow and eventually become
important when considering sufficiently long periods of time (of the
order of the evaporation time), but well before the Planckian regime
is reached. In addition, the assumption of a simple correlation
between the fluctuations of the energy flux crossing the horizon and
far from it, which was made in earlier work on spherically-symmetric
induced fluctuations, is carefully scrutinized and found to be
invalid. Our analysis suggests the existence of an infinite amplitude
for the fluctuations when trying to localize the horizon as a
three-dimensional hypersurface, as in the classical case, and, as a
consequence, a more accurate picture of the horizon as possessing a
finite effective width due to quantum fluctuations. This is supported
by a systematic analysis of the noise kernel in curved spacetime
smeared with different functions under different conditions, the
details are collected in the appendices. This case study shows a
pathway for probing quantum metric fluctuations near the horizon and
understanding their physical meaning.
\end{abstract}


\maketitle

\section{Introduction}
\label{sec1}

Studying the dynamics of quantum fields in a fixed curved spacetime,
Hawking found that black holes emit thermal radiation with a
temperature inversely proportional to their mass \cite{hawking75}.
When the back reaction of the quantum fields on the spacetime dynamics
is included, one expects that the mass of the black hole decreases as
thermal radiation at higher and higher temperatures is emitted. This
picture, which constitutes the process known as black hole
evaporation, is indeed obtained from semiclassical gravity
calculations which are believed to be valid at least before the
Planckian scale is reached \cite{bardeen81,massar95}.

Semiclassical gravity \cite{birrell94,wald94,flanagan96} is a mean
field description that neglects the fluctuations of the spacetime
geometry. However, a number of studies have suggested the existence
of large fluctuations near black hole horizons
\cite{sorkin95,sorkin97,casher97,marolf03} (and even instabilities
\cite{mazur04}) with characteristic time-scales much shorter than the
black hole evaporation time. In all of them\footnote{At least those
which provide a relativistic description. The argument in
Refs.~\cite{sorkin95,sorkin97} is based on a non-relativistic
description and it is not obvious how to make some of our statements
precise in that case. However, a natural generalization to the
relativistic case is provided in Ref.~\cite{marolf03}, which does
fall into this category.} either states which are singular on the
horizon (such as the Boulware vacuum for Schwarzschild spacetime)
were explicitly considered, or fluctuations were computed with
respect to those states and found to be large near the horizon.
Whether these huge fluctuations are of a generic nature or an
artifact arising from the consideration of states singular on the
horizon is an issue that deserves further investigation.  By
contrast, the fluctuations for states regular on the horizon were
estimated in Ref.~\cite{wu99} and found to be small even when
integrated over a time of the order of the evaporation time.

These apparently contradictory claims and the fact that most claims on
black hole horizon fluctuations were based on qualitative arguments
and/or semi-quantitative estimates prompted us here to strive for a
more quantitative and self-consistent description\footnote{Previous
attempts on this problem with similar emphasis by Raval, Sinha and
one of us have appeared in Refs.~\cite{hu99b,hu03b}. The apparent
difference between the conclusions in Ref.~\cite{hu99b} and what is
reported here will be explained below.}.
For this endeavor we follow the stochastic gravity program
\cite{calzetta94,martin99a,martin99b,hu03a,hu04a}. We will consider
the fluctuations of metric perturbations around a black hole geometry
interacting with a quantum scalar field whose stress tensor drives the
dynamics. The evolution of the mean background geometry is given by
the semiclassical Einstein equation (with self-consistent back
reaction from the expectation value of the stress tensor) while the
metric fluctuations obey an Einstein-Langevin equation
\cite{hu95a,hu95b,campos96,lombardo97} with a Gaussian stochastic
source whose correlation function is given by the noise kernel, which
characterizes the fluctuations of the stress tensor of the quantum
fields. In contrast to the claims made before, we find here that even
for states regular on the horizon the accumulated fluctuations become
significant by the time the black hole mass has changed substantially,
but well before reaching the Planckian regime. Our result is different
from those obtained in prior studies, but in agreement with earlier
work by Bekenstein~\cite{bekenstein84}.

The stochastic gravity program provides perhaps the best available
framework to study quantum metric fluctuations, because while
semiclassical gravity is a mean field description that does not take
into account quantum metric fluctuations, the Einstein-Langevin
equation enables one to solve for the dynamics of metric fluctuations
induced by the fluctuations of the stress tensor of the quantum
fields.  Furthermore, the correlation functions that one obtains are
equivalent to the quantum correlation functions for the metric
perturbations around the semiclassical background that would follow
from a quantum field theory treatment, up to a given order in an
expansion in terms of the inverse number of fields
\cite{roura03b,hu04b}. The quantization of these metric perturbations
should be understood in the framework of a low-energy effective field
theory approach to quantum gravity \cite{burgess04}, which is expected
to provide reliable results for phenomena involving typical
length-scales much larger than the Planck length even if the
microscopic details of the theory at Planckian scales are not
known.\footnote{This approach has been mainly applied to weak field
situations, such as the study of quantum corrections to the Newtonian
potential for particles in a Minkowski background
\cite{donoghue94,donoghue97}. However, it is particularly interesting
to apply it also to strong field situations involving cosmological
\cite{weinberg05} or black hole spacetimes.}

A crucial relation assumed in previous investigations
\cite{bekenstein84,wu99}\footnote{See, however,
Refs.~\cite{parentani01b,parentani02}, where those correlators were
shown to vanish in an effectively two-dimensional model.} of the
problem of metric fluctuations driven by quantum matter field
fluctuations of states regular on the horizon (as far as the
expectation value of the stress tensor is concerned) is the existence
of correlations between the outgoing energy flux far from the horizon
and a negative energy flux crossing the horizon, based on energy
conservation arguments. Using semiclassical gravity, such correlations
have been confirmed for the expectation value of the energy fluxes,
provided that the mass of the black hole is much larger than the
Planck mass. However, a more careful analysis, summarized in
Sec.~\ref{sec4}, shows that no such simple connection exists for
energy flux fluctuations. It also reveals that the fluctuations on the
horizon are in fact divergent unless it is treated as an object with a
finite width rather than a three-dimensional hypersurface, as in the
classical case, and one needs to find an appropriate way of probing
the metric fluctuations near the horizon and extracting physically
meaningful information. This is a new challenge in the study of metric
fluctuations which demands some careful thoughts on what they mean
physically and how they can be probed operationally. In
Appendices~\ref{appA} and \ref{appB} we give a systematic analysis of
the noise kernel in curved spacetime smeared with different functions
under different conditions.  The non-existence of this commonly
invoked relation in this whole subject matter illustrates the
limitations of heuristic arguments and the necessity of a detailed and
consistent formalism to study the fluctuations near the horizon, in
terms of their magnitude, how they are measured and their
consequences.

A few technical remarks are in order to delimit the problem under
study: First, we will restrict our attention to the
spherically-symmetric sector of metric fluctuations, which necessarily
implies a partial description of the fluctuations. That is because,
contrary to the case for semiclassical gravity solutions, even if one
starts with spherically-symmetric initial conditions, the stress
tensor fluctuations will induce fluctuations involving higher
multipoles. Thus, the multipole structure of the fluctuations is far
richer than that of spherically-symmetric semiclassical gravity
solutions, but this also means that obtaining a complete solution
(including all multipoles) for fluctuations rather than the mean value
is much more difficult.

Second, for black hole masses much larger than the Planck mass
(otherwise the effective field theory description will break down
anyway), one can introduce a useful adiabatic approximation involving
inverse powers of the black hole mass. To obtain results to lowest
order, it is sufficient to compute the expectation value of the stress
tensor operator and its correlation functions in Schwarzschild
spacetime. The corrections, proportional to higher powers of the
inverse mass, can be neglected for sufficiently massive black holes.

Third, when studying the dynamics of induced metric fluctuations, the
additional contribution to the stress tensor expectation value which
results from evaluating it using the perturbed metric is often
neglected. In the consideration of fluctuations for an evaporating
black hole such a term (which will be denoted by $\langle
\hat{T}_{ab}^{(1)} [g+h] \rangle_\mathrm{ren}$ in Sec.~\ref{sec3})
becomes important when it builds up for long times. The importance of
this term is clear when comparing with the simple estimate made by Wu
and Ford in Ref.~\cite{wu99}, where $\langle \hat{T}_{ab}^{(1)} [g+h]
\rangle _\mathrm{ren}$ was neglected and the fluctuations were found
to be small even when integrated over long times, of the order of the
evaporation time of the black hole.

The paper is organized as follows. In Sec.~\ref{sec2} we briefly
review the results for the evolution of the mean field geometry of an
evaporating black hole obtained in the context of semiclassical
gravity. The framework of stochastic gravity is then applied in
Sec.~\ref{sec3} to the study of the spherically-symmetric sector of
fluctuations around the semiclassical gravity solution for an
evaporating black hole. It has been previously assumed that an exact
correlation between the fluctuations of the negative energy flux
crossing the horizon and the flux far from it exists. In this paper we
want to question this assumption, but in the presentation in
Sec.~\ref{sec3} we accept temporarily such a working hypothesis just
so that we can have the common ground to compare our results with
those in the literature. In Sec.~\ref{sec4}, we present a careful
analysis of this assumption, and show that this supposition is
invalid. Further details of this proof can be found in the
Appendices~\ref{appA} and \ref{appB} . Finally, in Sec.~\ref{sec5} we
discuss several implications of our results and suggest some
directions for further investigation.

Throughout the paper we use Planckian units with $\hbar=c=G=1$ and the
$(+,+,+)$ convention of Ref.~\cite{misner73}. We also make use of the
abstract index notation of Ref.~\cite{wald84}. Latin indices denote
abstract indices, whereas Greek indices are employed whenever a
particular coordinate system is considered.

\section{Mean evolution of an evaporating black hole}
\label{sec2}

\emph{Semiclassical gravity} provides a mean field description of the
dynamics of a classical spacetime where the gravitational
back reaction of quantum matter fields is included self-consistently
\cite{birrell94,wald94,flanagan96}. It is believed
to be applicable to situations involving length-scales much larger
than the Planck scale and for which the quantum back-reaction effects
due to the metric itself can be neglected as compared to those due to
the matter fields.
The dynamics of the metric $g_{ab}$ is governed by the
semiclassical Einstein equation:
\begin{equation}
G_{ab} \left[ g \right] = \kappa \left\langle \hat{T}_{ab}
[g] \right\rangle _\mathrm{ren} \label{einstein1},
\end{equation}
where $\langle \hat{T}_{ab} [g] \rangle _\mathrm{ren}$ is
the renormalized expectation value of the stress tensor operator of
the quantum matter fields and $\kappa = 8 \pi / m_\mathrm{p}^2$ with
$m_\mathrm{p}^2$ being the Planck mass. One must solve both the
semiclassical Einstein equation and the equation of motion for the
matter fields evolving in that geometry, whose solution is needed to
evaluate $\langle \hat{T}_{ab} [g] \rangle _\mathrm{ren}$
self-consistently.

An important application of semiclassical gravity is the study of
black hole evaporation due to the back reaction of the Hawking
radiation emitted by the black hole on the spacetime geometry.
This has been studied in some detail for spherically symmetric black
holes \cite{bardeen81,massar95}. For a general spherically-symmetric
metric there always exists a system of coordinates in which it takes
the form
\begin{equation}
ds^2 = - e^{2 \psi(v,r)} ( 1 - 2 m(v,r)/r ) dv^2
+ 2 e^{\psi(v,r)} dv dr
+ r^2 \left( d\theta^2 + \sin^2 \theta d\varphi^2 \right)
\label{metric1}.
\end{equation}
This completely fixes the gauge freedom under local diffeomorphism
transformations except for an arbitrary function of $v$ that can be
added to the function $\psi(v,r)$ and is related to the freedom in
reparametrizing $v$ (we will see below how this can also be fixed).
In general this metric exhibits an \emph{apparent horizon}, where the
expansion of the outgoing radial null geodesics vanishes and which
separates regions with positive and negative expansion for those
geodesics, at those radii that correspond to (odd degree) zeroes of
the $vv$ metric component. We denote the location of the apparent
horizon by $r_\mathrm{AH}(v)=2M(v)$, where $M(v)$ satisfies the
equation $2m(2M(v),v)=2M(v)$.

Spherical symmetry implies that the components $T_{\theta r}$,
$T_{\theta v}$, $T_{\varphi r}$ and $T_{\varphi v}$ vanish and the
remaining components are independent of the angular coordinates.
Under these conditions the various components of the semiclassical
Einstein equation associated with the metric in Eq.~(\ref{metric1})
become
\begin{eqnarray}
\frac{\partial m}{\partial v} &=& 4 \pi r^2 T_v^r
\label{einstein2a},\\
\frac{\partial m}{\partial r} &=& - 4 \pi r^2 T_v^v
\label{einstein2b},\\
\frac{\partial \psi}{\partial r} &=& 4 \pi r T_{rr}
\label{einstein2c},
\end{eqnarray}
where in the above and henceforth we simply use $T_{\mu \nu}$ to
denote the expectation value $\langle \hat{T}_{\mu \nu} [g] \rangle
_\mathrm{ren}$ and employ Planckian units (with $m_\mathrm{p}^2=1$).
Note that the arbitrariness in $\psi$ can be eliminated by choosing a
parametrization of $v$ such that $\psi$ takes a particular value at a
given radius (we will choose that it vanishes at $r=2M(v)$, where the
apparent horizon is located); $\psi$ is then entirely fixed by
Eq.~(\ref{einstein2c}).

Solving Eqs.~(\ref{einstein2a})-(\ref{einstein2c}) is no easy task.
However, one can introduce a useful adiabatic approximation in the
regime where the mass of the black hole is much larger than the Planck
mass, which is in any case a necessary condition for the semiclassical
treatment to be valid. What this entails is that when $M \gg 1$
(remember that we are using Planckian units) for each value of $v$ one
can simply substitute $T_{\mu \nu}$ by its ``parametric value'' -- by
this we mean the expectation value of the stress energy tensor of the
quantum field in a Schwarzschild black hole with a mass corresponding
to $M(v)$ evaluated at that value of $v$. This is in contrast to its
dynamical value, which should be determined by solving
self-consistently the semiclassical Einstein equation for the
spacetime metric and the equations of motion for the quantum matter
fields.
This kind of approximation introduces errors of higher order in
$L_\mathrm{H} \equiv B/M^2$ ($B$ is a dimensionless parameter that
depends on the number of massless fields and their spins and accounts
for their corresponding grey-body factors; it has been estimated to be
of order $10^{-4}$ \cite{page76}), which are very small for black
holes well above Planckian scales. These errors are due to the fact
that $M(v)$ is not constant and that, even for a constant $M(v)$, the
resulting static geometry is not exactly Schwarzschild because the
vacuum polarization of the quantum fields gives rise to a
non-vanishing $\langle \hat{T}_{ab} [g] \rangle _\mathrm{ren}$
\cite{york85}.

The expectation value of the stress tensor for Schwarzschild spacetime
has been found to correspond to a thermal flux of radiation (with
$T_v^r = L_\mathrm{H} / (4 \pi r^2)$) for large radii and of order
$L_\mathrm{H}$ near the horizon\footnote{The natural quantum state for
a black hole formed by gravitational collapse is the Unruh vacuum,
which corresponds to the absence of incoming radiation far from the
horizon. The expectation value of the stress tensor operator for that
state is finite on the future horizon of Schwarzschild, which is the
relevant one when identifying a region of the Schwarzschild geometry
with the spacetime outside the collapsing matter for a black hole
formed by gravitational collapse.}
\cite{candelas80,page82,howard84a,howard84b,anderson95}. This shows
the consistency of the adiabatic approximation for $L_\mathrm{H} \ll
1$: the right-hand side of Eqs.~(\ref{einstein2a})-(\ref{einstein2c})
contains terms of order $L_\mathrm{H}$ and higher, so that the
derivatives of $m(v,r)$ and $\psi(v,r)$ are indeed small.
Furthermore, one can use the $v$ component of the stress-energy
conservation equation
\begin{equation}
\frac{\partial \left( r^2 T_v^r \right)}{\partial r}
+ r^2 \frac{\partial T_v^v}{\partial v} = 0
\label{conservation1},
\end{equation}
to relate the $T^r_v$ components on the horizon and far from it.
Integrating Eq.~(\ref{conservation1}) radially, one gets
\begin{equation}
(r^2 T^r_v) (r=2M(v),v) = (r^2 T^r_v) (r \approx 6M(v),v) +
O(L_{\mathrm{H}}^2), \label{conservation2}
\end{equation}
where we considered a radius sufficiently far from the horizon, but
not arbitrarily far (\emph{i.e.} $2M(v) \ll r \ll
M(v)/L_\mathrm{H}$). The second condition is necessary to ensure that
the size of the horizon has not changed much since the value of $v'$
at which the radiation crossing the sphere of radius $r$ at time $v$
left the region close to the horizon. Note that while in the nearly
flat region (for large radii) $T_v^r$ corresponds to minus the
outgoing energy flux crossing the sphere of radius $r$, on the
horizon, where $ds^2 = 2 e^{\psi(v,r)} dv dr + r^2 \left( d\theta^2 +
\sin^2 \theta d\varphi^2 \right)$, $T_v^r$ equals $T_{vv}$, which
corresponds to the null energy flux crossing the horizon. Hence,
Eq.~(\ref{conservation2}) relates the positive energy flux radiated
away far from the horizon and the negative energy flux crossing the
horizon. Taking into account this connection between energy fluxes and
evaluating Eq.~(\ref{einstein2a}) on the apparent horizon, we finally
get the equation governing the evolution of its size:
\begin{equation}
\frac{d M}{d v} = - \frac{B}{M^2}
\label{einstein3}.
\end{equation}
Unless $M(v)$ is constant, the event horizon and the apparent horizon
do not coincide. However, in the adiabatic regime their radii are
related, differing by a quantity of higher order in $L_\mathrm{H}$:
$r_\mathrm{EH}(v) = r_\mathrm{AH}(v) \, (1 + O(L_\mathrm{H}))$.

We close this section with an explanation of why we did not have to
deal with terms involving higher-order derivatives and even non-local
terms when considering the expectation value of the stress tensor as
one would expect for geometries with sufficiently arbitrary spacetime
dependence of certain metric components. (This can be seen in explicit
calculations for arbitrary Robertson-Walker geometries
\cite{calzetta94,calzetta97c} or arbitrary small metric perturbations
around specific backgrounds \cite{calzetta87,campos94}.) In our case
such non-local and higher-order derivative terms would also appear in
the exact expression of the stress tensor expectation value for
arbitrary $m(v,r)$ and $\psi(v,r)$ functions. However, the adiabatic
approximation for $M \gg 1$ that we have employed effectively gets rid
of them since one can replace the higher-derivative terms using
Eqs.~(\ref{einstein2a})-(\ref{einstein2c}) recursively and taking into
account that the terms on the right-hand side are of order
$1/M^2$. Therefore, higher-order derivative terms correspond to higher
powers of $1/M^2$ and are highly suppressed for $M \gg 1$.
Note that this argument, which is based on the black hole size being
much larger than the Planck length, is closely related to the order
reduction prescription \cite{parker93,flanagan96} that is often used
to deal with higher-derivative terms in semiclassical gravity and
other back-reaction problems.

\section{Spherically-symmetric induced fluctuations}
\label{sec3}

There are situations in which the fluctuations of the stress tensor
operator and the metric fluctuations that they induce may be
important, so that the mean field description provided by
semiclassical gravity is incomplete and even fails to capture the
most relevant phenomena (the generation of primordial cosmological
perturbations constitutes a clear example of that). \emph{Stochastic
gravity} \cite{calzetta94,martin99a,martin99b,hu03a,hu04a} provides
a framework to study those fluctuations. Its centerpiece is the
Einstein-Langevin equation \cite{hu95a,hu95b,campos96,lombardo97}
\begin{equation}
G_{ab}^{(1)}\left[ g+h\right] =\kappa \left\langle
\hat{T}_{ab}^{(1)} [g+h] \right\rangle _\mathrm{ren}
+\kappa \, \xi_{ab}\left[ g\right]
\label{einst-lang1},
\end{equation}
which governs the dynamics of the metric fluctuations around a
background metric $g_{ab}$ that corresponds to a given solution of
semiclassical gravity. The superindex $(1)$ indicates that only the
terms linear in the metric perturbations should be considered, and
$\xi_{ab}$ is a Gaussian stochastic source with vanishing expectation
value and correlation function\footnote{Throughout the paper we use
the notation $\langle \ldots \rangle_\xi$ for stochastic averages over
all possible realizations of the source $\xi_{ab}$ to distinguish them
from quantum averages, which are denoted by $\langle \ldots \rangle$.}
$\langle \xi_{ab} (x) \xi_{cd} (x') \rangle_\xi = (1/2) \langle \{
\hat{t}_{ab} (x), \hat{t}_{cd} (x') \} \rangle$ (with $\hat{t}_{ab}
\equiv \hat{T}_{ab} - \langle \hat{T}_{ab} \rangle$), where the term
on the right-hand side, which accounts for the stress tensor
fluctuations within this Gaussian approximation, is commonly known as
the noise kernel and denoted by $N_{abcd}(x,x')$. In this framework
the metric perturbations are still classical but
stochastic. Nevertheless, one can show that the correlation functions
for the metric perturbations that one obtains in stochastic gravity
are equivalent through order $1/N$ to the quantum correlation
functions that would follow from a quantum field theory treatment when
considering a large number of fields $N$ \cite{roura03b,hu04b}. In
particular, the symmetrized two-point function consists of two
contributions: \emph{intrinsic} and \emph{induced} fluctuations. The
intrinsic fluctuations are a consequence of the quantum width of the
initial state of the metric perturbations, and they are obtained in
stochastic gravity by averaging over the initial conditions for the
solutions of the homogeneous part of Eq.~(\ref{einst-lang1})
distributed according to
the reduced Wigner function associated with the initial quantum state
of the metric perturbations. On the other hand, the induced
fluctuations are due to the quantum fluctuations of the matter fields
interacting with the metric perturbations, and they are obtained by
solving the Einstein-Langevin equation using a retarded propagator
with vanishing initial conditions.

In this section we study the spherically-symmetric sector
[\emph{i.e.}, the monopole contribution, which corresponds to $l=0$,
in a multipole expansion in terms of spherical harmonics
$Y_{lm}(\theta,\phi)$] of metric fluctuations for an evaporating black
hole. 
In this case only induced fluctuations are possible. The fact that
intrinsic fluctuations cannot exist can be clearly seen if one
neglects vacuum polarization effects, since Birkhoff's theorem forbids
the existence of spherically-symmetric free metric perturbations in
the exterior vacuum region of a spherically-symmetric black hole that
keep the ADM mass constant. Even when vacuum polarization effects are
included, spherically-symmetric perturbations, characterized by
$m(v,r)$ and $\psi(v,r)$, are not independent degrees of freedom. This
follows from Eqs.~(\ref{einstein2a})-(\ref{einstein2c}), which can be
regarded as constraint equations.

The fluctuations of the stress tensor are inhomogeneous and
non-spherically-symmetric even if we choose a spherically-symmetric
vacuum state for the matter fields (spherical symmetry simply implies
that the angular dependence of the noise kernel in spherical
coordinates is entirely given by the relative angle between the
spacetime points $x$ and $x'$). This means that, in contrast to the
semiclassical gravity case, projecting onto the $l = 0$ sector of
metric perturbations does not give an exact solution of the
Einstein-Langevin equation in the stochastic gravity approach that we
have adopted here. Nevertheless, restricting to spherical symmetry in
this way gives more accurate results than two-dimensional
dilaton-gravity models resulting from simple dimensional reduction
\cite{trivedi93,strominger93,lombardo99}. This is because we project
the solutions of the Einstein-Langevin equation just at the end,
rather than considering only the contribution of the $s$-wave modes to
the classical action for both the metric and the matter fields from
the very beginning. Hence, an infinite number of modes for the matter
fields with $l \neq 0$ contribute to the $l = 0$ projection of the
noise kernel, whereas only the $s$-wave modes for each matter field
would contribute to the noise kernel if dimensional reduction had been
imposed right from the start, as done in
Refs.~\cite{parentani01a,parentani01b,parentani02} as well as in
studies of two-dimensional dilaton-gravity models.

The Einstein-Langevin equation for the spherically-symmetric sector of
metric perturbations can be obtained by considering linear
perturbations of $m(v,r)$ and $\psi(v,r)$, projecting the stochastic
source that accounts for the stress tensor fluctuations to the $l=0$
sector, and adding it to the right-hand side of
Eqs.~(\ref{einstein2a})-(\ref{einstein2c}). We will focus our
attention on the equation for the evolution of $\eta(v,r)$, the
perturbation of $m(v,r)$:
\begin{equation}
\frac{\partial (m + \eta)}{\partial v} = - \frac{B}{(m + \eta)^2}
+ 4 \pi r^2 \xi_v^r + O \left(L_\mathrm{H}^2 \right)
\label{einst-lang2},
\end{equation}
which reduces, after neglecting terms of order $L_\mathrm{H}^2$ or
higher, to the following equation to linear order in $\eta$:
\begin{equation}
\frac{\partial \eta}{\partial v} = \frac{2 B}{m^3} \eta
+ 4 \pi r^2 \xi_v^r
\label{einst-lang3}.
\end{equation}
It is important to emphasize that in Eq.~(\ref{einst-lang2}) we
assumed that the change in time of $\eta(v,r)$ is sufficiently slow so
that the adiabatic approximation employed in the previous section to
obtain the mean evolution of $m(v,r)$ can also be applied to the
perturbed quantity $m(v,r)+\eta(v,r)$. This is guaranteed as long as
the term corresponding to the stochastic source is of order
$L_\mathrm{H}$ or higher, a point that will be discussed below.

Obtaining the noise kernel which determines the correlation function
for the stochastic source is highly nontrivial even if we compute it
on the Schwarzschild spacetime, which is justified in the adiabatic
regime for the background geometry. As implicitly done in prior work
(for instance in Refs.~\cite{bekenstein84,wu99}; see, however,
Refs.~\cite{parentani01b,parentani02}), we will assume in this section
that the fluctuations of the radiated energy flux far from the horizon
are exactly correlated with the fluctuations of the negative energy
flux crossing the horizon. This is a crucial assumption which implies
an enormous simplification and allows a direct comparison with the
results in the existing literature, and its validity will be analyzed
more carefully in the next section.\footnote{This simple relation
between the energy flux crossing the horizon and the flux far from it
is valid for the expectation value of the stress tensor, which is
based on an energy conservation argument for the $T_v^r$ component.
In most of the literature this relation is assumed to hold also for
fluctuations. However, in the next section we will show that this is
an incorrect assumption. Therefore, results derived from this
assumption and conclusions drawn are in principle suspect. (This
misstep is understandable because most authors have not acquired as
much insight into the nature of fluctuations phenomena as now.) Our
investigation testifies to the necessity of a complete reexamination
of all cases afresh. In fact, an evaluation of the noise kernel near
the horizon seems unavoidable for the consideration of fluctuations
and back-reaction issues.}

Since the generation of Hawking radiation is especially sensitive to
what happens near the horizon, from now on we will concentrate on the
metric perturbations near the horizon\footnote{This means that
possible effects on the Hawking radiation due to the fluctuations of
the potential barrier for the radial mode functions will be missed by
our analysis.} and consider $\eta(v) = \eta(v,2M(v))$. Assuming that
the fluctuations of the energy flux crossing the horizon and those far
from it are exactly correlated, from Eq.~(\ref{einst-lang3}) we have
\begin{equation}
\frac{d \eta(v)}{d v} = \frac{2 B}{M^3(v)} \eta(v)
+ \xi(v)
\label{einst-lang4},
\end{equation}
where $\xi(v) \equiv (4 \pi r^2\, \xi_v^r) (v,r \approx 6M(v))$. The
correlation function for the spherically-symmetric fluctuation
$\xi(v)$ is determined by the integral over the whole solid angle of
the $N^{r\;r}_{\;v\;v}$ component of the noise kernel, which is given
by $(1/2) \langle \{ \hat{t}_v^r (x), \hat{t}_v^r (x') \} \rangle$.
The $l=0$ fluctuations of the energy flux of Hawking radiation far
from a black hole formed by gravitational collapse, characterized also
by $(1/2) \langle \{ \hat{t}_v^r (x), \hat{t}_v^r (x') \} \rangle$
averaged over the whole solid angle, have been studied in
Ref.~\cite{wu99}. Its main features are a correlation time of order
$M$ and a characteristic fluctuation amplitude of order $\epsilon_0 /
M^4$ (this is the result of smearing the stress tensor two-point
function, which diverges in the coincidence limit, over a period of
time of the order of the correlation time). The order of magnitude of
$\epsilon_0$ has been estimated to lie between $0.1 B$ and $B$
\cite{bekenstein84,wu99}.  For simplicity, we will consider quantities
smeared over a time of order $M$. We can then introduce the Markovian
approximation $(\epsilon_0 / M^3(v)) \delta(v-v')$, which
coarse-grains the information on features corresponding to time-scales
shorter than the correlation time $M$. Under those conditions $r^2
\xi^r_v$ is of order $1/M^2$ and thus the adiabatic approximation made
when deriving Eq.~(\ref{einst-lang2}) is justified.

The stochastic equation (\ref{einst-lang4}) can be solved in the usual
way and the correlation function for $\eta(v)$ can then be computed.
Alternatively, one can follow Bekenstein \cite{bekenstein84} and
derive directly an equation for $\langle \eta^2 (v) \rangle_\xi$. This
is easily done multiplying Eq.~(\ref{einst-lang4}) by $\eta(v)$ and
taking the expectation value. The result is
\begin{equation}
\frac{d}{dv} \langle \eta^2 (v) \rangle_\xi
= \frac{4 B}{M^3(v)} \langle \eta^2 (v) \rangle_\xi
+ 2 \langle \eta (v) \xi (v) \rangle_\xi
\label{fluct1}.
\end{equation}
For delta-correlated noise (the Stratonovich prescription is the
appropriate one in this case), $\langle \eta (v) \xi (v) \rangle_\xi$
equals one half the time-dependent coefficient multiplying the delta
function $\delta (v-v')$ in the expression for $\langle \xi (v) \xi
(v') \rangle_\xi$, which is given by $\epsilon_0 / M^3(v)$ in our
case. Taking that into account, Eq.~(\ref{fluct1}) becomes
\begin{equation}
\frac{d}{dv} \langle \eta^2 (v) \rangle_\xi
= \frac{4 B}{M^3(v)}
\langle \eta^2 (v) \rangle_\xi + \frac{\epsilon_0}{M^3(v)}
\label{fluct2}.
\end{equation}
Finally, it is convenient to change from the $v$ coordinate to the
mass function $M(v)$ for the background solution. Eq.~(\ref{fluct2})
can then be rewritten as
\begin{equation}
\frac{d}{dM} \langle \eta^2 (M) \rangle_\xi
= - \frac{4}{M} \langle \eta^2 (M) \rangle_\xi
- \frac{(\epsilon_0 / B)}{M}
\label{fluct3}.
\end{equation}
The solutions of this equation are given by
\begin{equation}
\langle \eta^2 (M) \rangle_\xi
= \langle \eta^2 (M_0) \rangle_\xi \left(\frac{M_0}{M}\right)^4
+\frac{\epsilon_0}{4 B} \left[\left(\frac{M_0}{M}\right)^4 - 1\right]
\label{fluct4}.
\end{equation}
Provided that the fluctuations at the initial time corresponding to
$M=M_0$ are negligible (much smaller than $\sqrt{\epsilon_0 / 4B} \sim
1$), the fluctuations become comparable to the background solution
when $M \sim M_0^{2/3}$. Note that fluctuations of the horizon radius
of order one in Planckian units do not correspond to Planck scale
physics because near the horizon $\Delta R = r - 2M$ corresponds to a
physical distance $L \sim \sqrt{M \, \Delta R}$, as can be seen
from the line element for Schwarzschild, $ds^2 = - (1-2M/r) dt^2 +
(1-2M/r)^{-1} dr^2 + r^2 (d\theta^2 + \sin^2 \theta d\varphi^2)$, by
considering pairs of points at constant $t$. So $\Delta R \sim 1$
corresponds to $L \sim \sqrt{M}$, whereas a physical distance of order
one is associated with $\Delta R \sim 1/M$, which corresponds to an
area change of order one for spheres with those radii. One can,
therefore, have initial fluctuations of the horizon radius of order
one for physical distances well above the Planck length provided that
we consider a black hole with a mass much larger than the Planck
mass. One expects that the fluctuations for states that are regular on
the horizon correspond to physical distances not much larger than the
Planck length, so that the horizon radius fluctuations would be much
smaller than one for sufficiently large black hole
masses. Nevertheless, that may not be the case when dealing with
states which are singular on the horizon, with estimated fluctuations
of order $M^{1/3}$ or even $\sqrt{M}$
\cite{casher97,marolf03,mazur04}. Confirming that the fluctuations are
indeed so small for regular states and verifying how generic, natural
and stable they are as compared to singular ones is a topic that we
plan to address in future investigations.

Our result for the growth of the fluctuations of the size of the black
hole horizon agrees with the result obtained by Bekenstein in
Ref.~\cite{bekenstein84} and implies that, for a sufficiently massive
black hole (with a few solar masses or a supermassive black hole), the
fluctuations become important before the Planckian regime is
reached. Strictly speaking, one cannot expect that a linear treatment
of the perturbations provides an accurate result when the fluctuations
become comparable to the mean value, but it signals a significant
growth of the fluctuations (at least until the nonlinear effects on
the perturbation dynamics become relevant).

This growth of the fluctuations which was found by Bekenstein and
confirmed here via the Einstein-Langevin equation seems to be in
conflict with the estimate given by Wu and Ford in
Ref.~\cite{wu99}. According to their estimate, the accumulated mass
fluctuations over a period of the order of the black hole evaporation
time ($\Delta t \sim M_0^3$) would be of the order of the Planck
mass. The discrepancy is due to the fact that the first term on the
right-hand side of Eq.~(\ref{einst-lang4}), which corresponds to the
perturbed expectation value $\langle \hat{T}_{ab}^{(1)} [g+h] \rangle
_\mathrm{ren}$ in Eq.~(\ref{einst-lang1}), was not taken into account
in Ref.~\cite{wu99}. The larger growth obtained here is a consequence
of the secular effect of that term, which builds up in time (slowly at
first, during most of the evaporation time, and becoming more
significant at late times when the mass has changed substantially) and
reflects the unstable nature of the background solution for an
evaporating black hole.\footnote{A clarification between our results
and the claims by Hu, Raval and Sinha in Ref.~\cite{hu99b} is in place
here: both use the stochastic gravity framework and perform an
analysis based on the Einstein-Langevin equation, so there should be
no discrepancy. However, the claim in Ref.~\cite{hu99b} was based on a
qualitative argument that focused on the stochastic source, but missed
the fact that the perturbations around the mean are unstable for an
evaporating black hole. Once this is taken into account, agreement
with the result obtained here is recovered.}

All this can be qualitatively understood as 
follows. Consider an evaporating black hole with initial mass $M_0$
and suppose that the initial mass is perturbed by an amount $\delta
M_0 = 1$. The mean evolution for the perturbed black hole (without
taking into account any fluctuations) leads to a mass perturbation
that grows like $\delta M = (M_0/M)^2 \, \delta M_0 = (M_0/M)^2$, so
that it becomes comparable to the unperturbed mass $M$ when $M \sim
M_0^{2/3}$, which coincides with the result obtained above. Such a
coincidence has a simple explanation: the fluctuations of the Hawking
flux slowly accumulated during most of the evaporating time, which
are of the order of the Planck mass, as found by Wu and Ford, give a
dispersion of that order for the mass distribution at the time when the
instability of the small perturbations around the background solution
start to become significant.

\section{Correlation between outgoing and ingoing energy fluxes}
\label{sec4}


In this section we address the issue whether the simple relation
between the energy flux crossing the horizon and the flux far from it
also holds for the fluctuations. As we show in Appendix~\ref{appC},
those correlations vanish for conformal fields in two-dimensional
spacetimes. (The correlation function for the outgoing and ingoing
null energy fluxes in an effectively two-dimensional model was
explicitly computed in Refs.~\cite{parentani01b,parentani02} and it
was indeed found to vanish.) On the other hand, in four dimensions the
correlation function does not vanish in general and correlations
between outgoing and ingoing fluxes do exist near the horizon (at
least partially). This point is also explained in Appendix~\ref{appC}.

For black hole masses much larger than the Planck mass, one can use
the adiabatic approximation for the background mean
evolution. Therefore, to lowest order in $L_\mathrm{H}$ one can
compute the fluctuations of the stress tensor in Schwarzschild
spacetime. In Schwarzschild, the amplitude of the fluctuations of $r^2
\hat{T}^r_v$ far from the horizon is of order $1/M^2$ ($= M^2 / M^4$)
when smearing over a correlation time of order $M$, which one can
estimate for a hot thermal plasma in flat space
\cite{campos98,campos99}
(see also Ref.~\cite{wu99} for a computation of the fluctuations of
$r^2 \hat{T}^r_v$ far from the horizon). The amplitude of the
fluctuations of $r^2 \hat{T}^r_v$ is thus of the same order as its
expectation value. However, their derivatives with respect to $v$ are
rather different: since the characteristic variation times
for the expectation value and the fluctuations are $M^3$ and $M$
respectively, $\partial (r^2 T^r_v) / \partial v$ is of order $1/M^5$
whereas $\partial (r^2 \xi^r_v) / \partial v$ is of order $1/M^3$.
This implies an additional contribution of order $L_\mathrm{H}$ due to
the second term in Eq.~(\ref{conservation1}) if one radially
integrates the same equation applied to stress tensor fluctuations
(the stochastic source in the Einstein-Langevin equation). Hence, in
contrast to the case of the mean value, the contribution from the
second term in Eq.~(\ref{conservation1}) cannot be neglected when
radially integrating since it is of the same order as the
contributions from the first term, and one can no longer obtain a
simple relation between the outgoing energy flux far from the horizon
and the energy flux crossing the horizon.

So far we have argued that the method employed for the mean value
cannot be employed for the fluctuations. Although one expects that
$r^2 \xi^r_v$ on the horizon and far from it will not be equal when
including the contributions that results from radially integrating the
second term in Eq.~(\ref{conservation1}), one might wonder whether
there is a possibility that those contributions would somehow cancel
out. That possibility can, however, be excluded using the following
argument. The smeared correlation function
\begin{equation}
\int dv h(v) \int dv' h(v')\,
r^4 \langle \xi^r_v (v,r) \xi^r_v (v',r) \rangle_\xi
\label{correlation1},
\end{equation}
where $h(v)$ is some appropriate smearing function and $\xi^r_v (v,r)$
has already been integrated over the whole solid angle, is divergent
on the horizon but finite far from it. Therefore, $r^2 \xi^r_v$ on the
horizon and far from it cannot be equal for each value of $v$.

Let us discuss in some more detail the fact that certain smearings of
the quantity $r^4 \langle \xi^r_v (v,r) \xi^r_v (v',r) \rangle_\xi$
are divergent on the horizon but finite far from it.
The smeared correlation function is related to the noise kernel as
follows:
\begin{equation}
\int dv dv' h(v) h(v')\,
r^4 \langle \xi^r_v (v,r) \xi^r_v (v',r) \rangle_\xi
= r^4 \int dv dv' h(v) h(v') \int d\Omega d\Omega'
N^{r\;r}_{\;v\;v} (v,r,\theta,\varphi;v',r,\theta',\varphi')
\label{correlation2}.
\end{equation}
The noise kernel is divergent in the coincident limit or for
null-separated points. Smearing the noise kernel along all directions
gives a finite result. However, although certain partial smearings
also give a finite result, others do not. For instance, smearing along
a timelike direction yields a finite result, whereas smearing on a
spacelike hypersurface yields in general a divergent result
\cite{ford05}. On the other hand, the result of smearing along two
``transverse'' null directions (two null directions sharing the same
orthogonal spacelike 2-surfaces) is also finite, but not for a
smearing along just one null direction even if we also smear along the
orthogonal spacelike directions. For $r>2M$ Eq.~(\ref{correlation2})
corresponds to a smearing along a timelike direction and gives a
finite result for the smeared correlation function, but on the horizon
it corresponds to a smearing along a single null direction and it is
divergent.

A proof of the results described in the previous paragraph is
provided in Appendices~\ref{appA} and \ref{appB} by considering a
product of smearing functions involving all directions and then taking
different kinds of limits in which the smearing size along certain
directions vanishes. One limit corresponds to taking a vanishing size
for the smearing along one of the null directions and we refer to it
as \emph{smearing along null geodesics}. The other corresponds to
taking vanishing sizes for certain spatial directions and we refer to
it as \emph{smearing along timelike curves}.  The proof proceeds in
two steps. First, in Appendix~\ref{appA} it is shown for the flat
space case. Then it is generalized to curved spacetimes in
Appendix~\ref{appB} using a quasilocal expansion in terms of Riemann
normal coordinates.

\section{Discussion}
\label{sec5}

Following the stochastic gravity program, in Sec.~\ref{sec3} we found
that the spherically-symmetric fluctuations of the horizon size of an
evaporating black hole become important at late times, and even
comparable to its mean value when $M \sim M_0^{2/3}$, where $M_0$ is
the mass of the black hole at some initial time when the fluctuations
of the horizon radius are much smaller than the Planck
length.\footnote{Remember that for large black hole masses this can
still correspond to physical distances much larger than the Planck
length, as explained in Sec.~\ref{sec3}.} This is consistent with the
result previously obtained by Bekenstein in Ref.~\cite{bekenstein84}.

It is important to realize that for a sufficiently massive black hole,
the fluctuations become significant well before the Planckian regime
is reached. More specifically, for a solar mass black hole they become
comparable to the mean value when the black hole radius is of the
order of $10 \mathrm{nm}$, whereas for a supermassive black hole with
$M \sim 10^7 M_\odot$, that happens when the radius reaches a size of
the order of $1 \mathrm{mm}$. One expects that in those circumstances
the low-energy effective field theory approach of stochastic gravity
should provide a reliable description.

It is worth mentioning that other properties of the black hole can
exhibit substantial fluctuations when taking into account the
back reaction of Hawking radiation. As pointed out by Page in
Ref.~\cite{page80}, by momentum conservation the fluctuations of the
total momentum of the Hawking radiation emitted will cause a recoil
of the position of the black hole, which will also fluctuate.
According to Page's estimate, the spread of the distribution for the
black hole position will become comparable to the size of the horizon
by the time an energy of order $M^{1/3}$ has been emitted. This kind
of fluctuations, which were not obtained in our calculation because
we restricted our attention to spherically-symmetric metric
perturbations, exhibit certain features that differ significantly
from those of our result. The fluctuations that we found take a much
longer time to build up and depend crucially on the unstable behavior
of small perturbations of the semiclassical solution, with
characteristic time-scales of the order of the evaporation time. On
the contrary, this unstable behavior plays no major role in the
growth of the position fluctuations. Furthermore, since this growth
is still much slower than the emission rate of Hawking quanta, in the
frame where the black hole is at rest the properties of the Hawking
radiation being emitted remain essentially unchanged when the
position fluctuations are taken into account, whereas the
fluctuations of the horizon size do imply fluctuations in the
temperature of the Hawking radiation.

Due to the nonlinear nature of the back-reaction equations, such as
Eq.~(\ref{einst-lang2}), the fact that the fluctuations of the horizon
size can grow and become comparable to the mean value implies
non-negligible corrections to the dynamics of the mean value
itself. This can be seen by expanding Eq.~(\ref{einst-lang2})
(evaluated on the horizon) in powers of $\eta$ and taking the
expectation value. Through order $\eta^2$ we get
\begin{eqnarray}
\frac{d (M(v) + \langle \eta(v) \rangle_\xi)}{d v}
&=& - \left \langle \frac{B}{(M(v) + \eta(v))^2} \right \rangle_\xi
\nonumber \\
&=& - \frac{B}{M^2(v)} \left[ 1 - \frac{2}{M(v)} \langle \eta (v) \rangle_\xi
+ \frac{3}{M^2(v)} \langle \eta^2 (v) \rangle_\xi + O \left( \frac{\eta^3}{M^3}
\right) \right]  \label{rad_correction}.
\end{eqnarray}
When the fluctuations become comparable to the mass itself, the
third term (and higher order terms) on the right-hand side is no
longer negligible and we get non-trivial corrections to
Eq.~(\ref{einstein3}) for the dynamics of the mean value. These
corrections can be interpreted as higher order radiative corrections
to semiclassical gravity that include the effects of metric
fluctuations on the evolution of the mean value. For instance, the
third term on the right-hand side of Eq.~(\ref{rad_correction})
would correspond to a two-loop Feynman diagram involving a matter
loop with an internal propagator for the metric perturbations
(restricted to the spherically-symmetric sector in our case), as
compared to just one matter loop, which is all that semiclassical
gravity can account for.

Does the existence of the significant deviations for the mean
evolution mentioned above imply that the results based on
semiclassical gravity obtained by Bardeen and Massar in
Refs.~\cite{bardeen81,massar95} are invalid? Several remarks are in
order. First of all, those deviations start to become significant only
after a period of the order of the evaporation time when the mass of
the black hole has decreased substantially. Secondly, since
fluctuations were not considered in those references, a direct
comparison cannot be established. However, we can compare the average
of the fluctuating ensemble with their results. Doing so exhibits an
evolution that deviates significantly when the fluctuations become
important. Nevertheless, if one considers a single member of the
ensemble at that time, its evolution will be accurately described by
the corresponding semiclassical gravity solution until the
fluctuations around that particular solution become important again,
after a period of the order of the evaporation time associated with
the new initial value of the mass at that time.

An interesting aspect that we have not addressed in this work, but
which is worth investigating, is the quantum coherence of those
fluctuations. It seems likely that, given the long time periods
involved and the size of the fluctuations, the entanglement between
the Hawking radiation emitted and the black hole spacetime geometry
will effectively decohere the large horizon fluctuations, rendering
them equivalent to an incoherent statistical ensemble.



In this paper we have taken a first step to put the study of metric
fluctuations in black hole spacetimes on a firmer basis by considering
a detailed derivation of the results from an appropriate formalism
rather than using heuristic arguments or simple estimates.  The spirit
is somewhat analogous to the study of the mean back-reaction effect of
Hawking radiation on a black hole spacetime geometry (both for black
holes in equilibrium and for evaporating ones) by considering the
solutions of semiclassical gravity in that case rather than just
relying on simple energy conservation arguments. In order to obtain an
explicit result from the stochastic gravity approach and compare with
earlier work, in Sec.~\ref{sec3} we employed a simplifying assumption
implicitly made in most prior work: the existence of a simple
connection between the outgoing energy flux fluctuations far from the
horizon and the negative energy flux fluctuations crossing the
horizon. In Sec.~\ref{sec4} we analyzed this assumption carefully and
showed it to be invalid. This strongly suggests that one needs to
study the stress tensor fluctuations from an explicit calculation of
the noise kernel near the horizon. This quantity is obtainable from
the stochastic gravity program and calculation is underway
\cite{phillips01,phillips03,eftekharzadeh07}.

A possible way to compute the noise kernel near the horizon could be
to use an approximation scheme based on a quasilocal expansion such as
Page's approximation \cite{page82} or similar methods corresponding to
higher order WKB expansions \cite{anderson95}.\footnote{Note, however,
that in most of these approaches the state of the quantum fields is
the Hartle-Hawking vacuum. For an evaporating black hole, one should
consider the Unruh vacuum.} With these techniques one can obtain an
approximate expression for the Wightman function of the matter fields,
which is the essential object needed to compute the noise
kernel. Unfortunately these approximations are only accurate for pairs
of points with a small separation scale and break down when it becomes
comparable to the black hole radius.  Therefore, especial care is
needed when studying the $l=0$ multipole since that corresponds to
averaging the noise kernel over the whole solid angle, which involves
typical separations for pairs of points on the horizon of the order of
the black hole radius, and one needs to make sure that the integral is
dominated by the contribution from small angular separations.
Alternatively, one might hope to gain some insight on the
fluctuations near a black hole horizon by studying the fluctuations
of the event horizon surrounding any geodesic observer in de Sitter
spacetime, which exhibits a number of similarities with the event
horizon of a black hole in equilibrium \cite{gibbons77a}. In
contrast to the black hole case, it may be possible to obtain exact
analytical results for de Sitter space due to its high degree of
symmetry.

Furthermore, as explained in Sec.~\ref{sec4} and shown in detail in
Appendices~\ref{appA} and \ref{appB}, the noise kernel smeared over
the horizon is divergent, and so are the induced metric
fluctuations. Hence, one cannot study the fluctuations of the horizon
as a three-dimensional hypersurface for each realization of the
stochastic source because the amplitude of the fluctuations is
infinite, even when restricting one's attention to the $l=0$
sector. Instead one should regard the horizon as possessing a finite
effective width due to quantum fluctuations. In order to characterize
its width one must find a sensible way of probing the metric
fluctuations near the horizon and extracting physically meaningful
information, such as their effect on the Hawking radiation emitted by
the black hole.  One possibility is to study how metric fluctuations
affect the propagation of a bundle of null geodesics
\cite{barrabes99,barrabes00,parentani01a,parentani01b,parentani02,ford97}.
One expects that this should provide a description of the effects on
the propagation of a test field whenever the geometrical optics
approximation is valid. However, preliminary analysis of simpler cases
with a quantum field theory treatment suggest that when including
quantum vacuum metric fluctuations the geometrical optics
approximation becomes invalid. Another possibility, which seems to
constitute a better probe of the metric fluctuations, is to analyze
their effect on the two-point quantum correlation functions of a test
field. The two-point functions characterize the response of a particle
detector for that field and can be used to obtain the expectation
value and the fluctuations of the stress tensor of the test field.

Finally, since the large fluctuations suggested in
Refs.~\cite{sorkin95,sorkin97,casher97,marolf03} involve time-scales
much shorter than the evaporation time (contrary to those considered
in this paper) and high multipoles, one expects that for a
sufficiently massive black hole the spacetime near the horizon can be
approximated by Rindler spacetime (identifying the black hole horizon
and the Rindler horizon) provided that we restrict ourselves to
sufficiently small angular scales. Thus, analyzing the effect of
including the interaction with the metric fluctuations on the
two-point functions of a test field propagating in flat space, which
is technically much simpler, could provide useful information for the
black hole case.

\begin{acknowledgments}
We thank Paul Anderson, Larry Ford, Valeri Frolov, Ted Jacobson, Don
Marolf, Emil Mottola, Don Page, Renaud Parentani and Rafael Sorkin
for useful discussions. B.~L.~H.\ appreciates the hospitality of Professor
Stephen Adler while visiting the Institute for Advanced Study,
Princeton in Spring 2007. This work is supported in part by an NSF
Grant PHY-0601550. A.~R.\ was also supported by LDRD funds from Los
Alamos National Laboratory.
\end{acknowledgments}

\appendix

\section{Correlations between outgoing and ingoing fluxes in
$1+1$ and $3+1$ dimensions}
\label{appC}

Any spacetime metric in $1+1$ dimensions is conformally flat (at least
locally) and can be written as
\begin{equation}
ds^2 = - C(u,v) du dv
\label{metric2}.
\end{equation}
In terms of these null coordinates, the conservation equation for the
stress tensor, $\nabla^a T_{ab} = 0$, reduces to
\begin{eqnarray}
\partial_v \hat{T}_{uu} + \partial_u \hat{T}_{vu} - \Gamma^u_{uu} T_{vu} &=& 0
\label{conservation2a}, \\
\partial_u \hat{T}_{vv} + \partial_v \hat{T}_{uv} - \Gamma^v_{vv} T_{uv} &=& 0
\label{conservation2b},
\end{eqnarray}
since all the other relevant Christoffel symbols vanish.
Taking into account that $\Gamma^u_{uu} = \partial_u (\ln C)$,
$\Gamma^v_{vv} = \partial_v (\ln C)$ and combining
Eqs.~(\ref{conservation2a})-(\ref{conservation2b}) we get
\begin{eqnarray}
\partial_v \hat{T}_{uu} + \partial_u \hat{T}_{vv}
&=& - C \, \partial_t ( \hat{T}_{uv} / C )
\label{conservation3a}, \\
\partial_v \hat{T}_{uu} - \partial_u \hat{T}_{vv}
&=& C \, \partial_x ( \hat{T}_{uv} / C )
\label{conservation3b},
\end{eqnarray}
where we introduced the coordinates $t = (u + v)/2$ and $x = (v -
u)/2$.

This result can be applied to the Schwarzschild geometry in $1+1$
dimensions identifying $t$ with the usual Killing time and $x$ with
the Regge-Wheeler coordinate $r_*$. In that case we have $C(t,r) =
(1-2M/r)$. If we consider a massless conformally coupled field in
$1+1$ dimensions (conformal and minimal coupling are equivalent in
that case), the trace of the stress tensor (which is related to the
$T_{uv}$ component) vanishes at the classical level and is entirely
given by the trace anomaly $\langle \hat{T}^\mu_\mu
\rangle_\mathrm{ren} = R / 24 \pi = M / 6 \pi r^3$. Since both the
trace anomaly and the conformal factor are time-independent, the term
on the right-hand side of Eqs.~(\ref{conservation3a})
vanishes. Therefore, it follows from Eq.~(\ref{conservation3a}) that
the generation of left-moving and right-moving null mean fluxes is
perfectly anticorrelated, which implies that the positive energy flux
of outgoing Hawking radiation equals, in absolute value, the negative
energy flux crossing the horizon. Moreover, from
Eq.~(\ref{conservation3a}) and the fact that $\langle \hat{T}_{uv}
\rangle_\mathrm{ren} / C = - \langle \hat{T}^\mu_\mu
\rangle_\mathrm{ren} / 4 = - M / 24 \pi r^3$, which implies
$C\, \partial_{r_*}\! (\langle \hat{T}_{uv} \rangle_\mathrm{ren} / C)
= C^2 \, M / 6 \pi r^4$, it is clear that the amount of anticorrelated
mean fluxes generated tends to zero for large radii and is largest at
$r=3M$.

For energy flux fluctuations, the situation is very different. The
trace anomaly does not fluctuate, \emph{i.e.}, the trace of the stress
tensor does not fluctuate for conformal fields
\cite{martin99a}. Hence, in $1+1$ dimensions neither correlated nor
anticorrelated fluctuations of the left-moving and right-moving null
fluxes can be generated in the absence of other interactions: both
fluxes are separately conserved.

On the other hand, although in $3+1$ dimensions one can try to use a
similar argument when considering sectors with certain symmetries, the
final conclusion is different.  For instance, if one uses in Minkowski
spacetime a coordinate system $\{u,v,y,z\}$ where $u=t-x$ and $v=t+x$
are null coordinates, the components $\hat{T}_{uy}$, $\hat{T}_{uz}$,
$\hat{T}_{vy}$, $\hat{T}_{vz}$ and $\hat{T}_{yz}$ vanish in the sector
which is rotationally invariant on the $yz$ plane. In that sector, the
$u$ and $v$ components of the conservation equation coincide with
those in the $1+1$ case, given by
Eqs.~(\ref{conservation2a})-(\ref{conservation2b}), with vanishing
Christoffel symbols. Therefore, the conditions for the generation of
correlated or anticorrelated null fluxes are still given by
Eqs.~(\ref{conservation3a})-(\ref{conservation3b}). However, in $3+1$
dimensions the $\hat{T}_{yy}$ and $\hat{T}_{zz}$ components also
contribute to the trace of the stress tensor. Hence, although the
trace does not fluctuate for conformal fields, $\hat{T}_{uv}$ does
fluctuate and its fluctuations coincide with those of
$(\hat{T}_{yy}+\hat{T}_{zz})/4$. Correlated and anticorrelated null
energy flux fluctuations can thus be generated.

The discussion in the previous paragraph can be extended to a general
spherically symmetric spacetime in the region where the gradient of
the radial coordinate is spacelike. This can be done by considering
angular coordinates for every sphere of constant radius and the
coordinates associated with the two radial null directions orthogonal
to them. Taking rotational symmetry into account, as done in the flat
space case, one can provide an argument similar to that in $1+1$
dimensions. However, as in the flat space case, despite the absence of
fluctuations in the trace of the stress tensor, $\hat{T}_{uv}$ will
still fluctuate due to the fluctuations of $(\hat{T}_{\theta\theta} +
\hat{T}_{\varphi\varphi} / \sin^2 \theta) / 4r^2$. Moreover, in this
case there will be additional contributions due to scattering off the
potential barrier (an effect that will be small near the horizon).

\section{Smearing of the noise kernel in flat space}
\label{appA}

In this appendix we will consider several kinds of smearings of the
noise kernel for the Minkowski vacuum. In Sec.~\ref{sec:null} we study
a product of smearing functions involving two null directions and the
two orthogonal spatial directions, and analyze the limit in which the
smearing size along one of the null directions vanishes, which is
shown to be divergent. On the other hand, in Sec.~\ref{sec:timelike} a
smearing along a timelike direction and the three orthogonal
directions is considered and it is shown that a finite smearing size
along the timelike direction is sufficient to have a finite result. In
contrast, in the limit of a vanishing smearing size along the timelike
direction the result is always divergent, even for non-vanishing
smearing sizes along all the spatial directions.

The noise kernel should be treated as distribution in spacetime
coordinates. It has a divergent coincidence limit and it involves
subtle integration prescriptions. However, the Fourier transforms of
this kind of distributions are much simpler to deal with. Therefore,
it is very convenient to perform our calculations in Fourier space and
we will do so throughout this appendix.

The noise kernel $N_{abcd}(x,x') = (1/2)\langle
\{\hat{T}_{ab}(x), \hat{T}_{cd}(x')\} \rangle - \langle
\hat{T}_{ab}(x) \rangle \langle \hat{T}_{cd}(x') \rangle$ for
a massless conformally coupled scalar field in the Minkowski vacuum
state has been obtained for instance in Ref.~\cite{martin00} and is
given in standard inertial coordinates by
\begin{equation}
N_{\mu\nu\rho\sigma}(x,x') = \frac{2}{3}
\left( 3 \mathcal{D}_{\mu (\rho} \mathcal{D}_{\sigma) \nu}
- \mathcal{D}_{\mu \nu}\mathcal{D}_{\rho \sigma} \right) N(x-x')
\label{noise1},
\end{equation}
where $\mathcal{D}_{\mu \nu} \equiv \eta_{\mu \nu} \Box -
\partial_\mu \partial_\nu$ and 
\begin{equation}
N(x-x') = \frac{1}{(1920 \pi)} \int \frac{d^4p}{(2 \pi)^4} \,
e^{i p (x-x')} \theta(-p^2)
\label{noise2}.
\end{equation}

\subsection{Smearing around a null geodesic}
\label{sec:null}

In this subsection we consider the case in which one smears around a
null geodesic.
Using the null coordinates $v=t+x$ and $u=t-x$, we define the smeared
version of the kernel $N(x-x')$ as
\begin{equation}
\mathcal{N} \equiv
\int du\, dv\, d^2 x\, g(u) h (v) f(\vec{x})
\int du' dv' d^2 x' g(u') h (v') f(\vec{x}')
N(v-v',u-u',\vec{x}-\vec{x}')
\label{smearing1},
\end{equation}
where we integrated the kernel with some smearing functions both for
the null coordinates $u$ and $v$, and for the orthogonal spatial
directions.  If we choose Gaussian smearing functions
\begin{eqnarray}
g(u) &=& (2 \pi \sigma_u^2)^{-\frac{1}{2}} \exp(-u^2 / 2 \sigma_u^2)
\label{gaussian1},\\
h(v) &=& (2 \pi \sigma_v^2)^{-\frac{1}{2}} \exp(-v^2 / 2 \sigma_v^2)
\label{gaussian3},\\
f(\vec{x})  &=& (2 \pi \sigma_r^2)^{-1} \exp(-\vec{x}^2 / 2 \sigma_r^2)
\label{gaussian2},
\end{eqnarray}
Eq.~(\ref{smearing1}) can be written as
\begin{equation}
\mathcal{N} =
\frac{1}{(2 \pi)^4 \sigma_v^2 \sigma_u^2 \sigma_r^4}
\int dV dU d^2 X e^{-\frac{V^2}{\sigma_v^2}} e^{-\frac{U^2}{\sigma_u^2}}
e^{- \frac{\vec{X}^2}{\sigma_r^2}}
\int d \Delta_v d \Delta_u d^2 \Delta \, e^{-\frac{\Delta_v^2}{4 \sigma_v^2}}
e^{-\frac{\Delta_u^2}{4 \sigma_u^2}} e^{-\frac{\vec{\Delta}^2}{4 \sigma_r^2}}
N(\Delta_v,\Delta_u,\vec{\Delta})
\label{smearing11},
\end{equation}
where we introduced the semisum and difference variables $U=(u+u')/2$,
$V=(v+v')/2$, $\vec{X}=(\vec{x}+\vec{x}')/2$, $\Delta_u=u'-u$,
$\Delta_v=v'-v$ and $\vec{\Delta}=\vec{x}'-\vec{x}$. The integrals for
$U$, $V$ and $\vec{X}$ can be readily performed and yield the
following result:
\begin{equation}
\mathcal{N} =
\frac{1}{(4 \pi)^{2} \sigma_v \sigma_u \sigma_r^2}
\int d \Delta_v d \Delta_u d^2 \Delta \, e^{-\frac{\Delta_v^2}{4 \sigma_v^2}}
e^{-\frac{\Delta_u^2}{4 \sigma_u^2}} e^{-\frac{\vec{\Delta}^2}{4 \sigma_r^2}}
N(\Delta_v,\Delta_u,\vec{\Delta})
\label{smearing12}.
\end{equation}
On the other hand, in order to compute the remaining integrals it is
convenient to work in Fourier space.  When considering the null
coordinates $v$ and $u$, it is useful to introduce the momenta
$p_v=(p_x-p_t)/2$ and $p_u=-(p_t+p_x)/2$ so that Eq.~(\ref{noise2})
becomes
\begin{equation}
N(x-x') = \frac{1}{(1920 \pi)} \int \frac{d^4p}{(2 \pi)^4} \,
e^{i [p_v (v-v') + p_u (u-u') + \vec{p} \cdot (\vec{x}-{x}')]}
\theta(p_v p_u - \vec{p}^{\, 2})
\label{noise3},
\end{equation}
where we used the vector notation for the transverse components
associated with the coordinates $y$ and $z$. Eq.~(\ref{smearing12})
can then be expressed in Fourier space as
\begin{equation}
\mathcal{N} =
\frac{1}{(1920 \pi)} \frac{1}{(2 \pi)^{4}} \int  d p_v d p_u d^2 p\,
e^{-p_v^2 \sigma_v^2} e^{-p_u^2 \sigma_u^2} e^{-\vec{p}^2 \sigma_r^2}\,
\theta(p_v p_u - \vec{p}^{\, 2})
\label{smearing13}.
\end{equation}
One can then infer that $\mathcal{N}$ diverges as $\sigma_u
\rightarrow 0$. This can be seen as follows. The integral in
Eq.~(\ref{smearing13}) gets two identical contributions from the
quadrants $(p_u,p_v>0)$ and $(p_u,p_v<0)$, whereas the remaining two
quadrants give a vanishing contribution. Moreover, since the integrand
is positive, the integral is also positive and greater than the same
integral restricted to a smaller domain of integration. Taking all
that into account, we have
\begin{eqnarray}
\mathcal{N} &\geq&
\frac{2}{(1920 \pi)(2 \pi)^{4}}
\int \limits_{\genfrac{}{}{0pt}{}{p_v p_u \geq \sigma_r^{-2}}{p_v,p_u>0}}
d p_u d p_v e^{-p_u^2 \sigma_u^2} e^{-p_v^2 \sigma_v^2}
\int_0^{\sigma_r^{-2}} d |\vec{p}|^2 e^{-1} \nonumber \\
&\geq& \frac{2 e^{-1}}{\sigma_r^2} \frac{1}{(1920 \pi)(2 \pi)^{4}}
\int_{2 \sigma_v/\sigma_r^{2}}^\infty d p_u e^{-p_u^2 \sigma_u^2}
\int_{\sigma_v^{-1}/2}^{\sigma_v^{-1}} d p_v e^{-1} \nonumber \\
&=& \frac{e^{-2}}{\sigma_r^2 \sigma_v} \frac{1}{(1920 \pi)(2 \pi)^{4}}
\int_{2 \sigma_v/\sigma_r^{2}}^\infty d p_u e^{-p_u^2 \sigma_u^2}
\sim \frac{1}{\sigma_u \sigma_v \sigma_r^2}
\label{smearing14}.
\end{eqnarray}
The last integral is divergent if one takes $\sigma_u \to 0$ (at least
for $\sigma_r \neq 0$ \footnote{The divergence of $\mathcal{N}$ as
$\sigma_u \rightarrow 0$ can also be proven for $\sigma_r=0$
(\emph{i.e.}, in the absence of smearing along the transverse spatial
directions) by taking $\sigma_r=0$ in Eq.~(\ref{smearing13}) and
replacing $\sigma_r^2$ with an arbitrary but fixed positive value in
Eq.~(\ref{smearing14}).}). Thus, $\mathcal{N}$ diverges unless
$\sigma_u \neq 0$.

Using the previous result, it is easy to discuss whether a smeared
version of the actual noise kernel (including the differential
operators), given by Eq.~(\ref{noise1}), also diverges when $\sigma_u
\to 0$. Each derivative in Eq.~(\ref{noise1}) gives rise to an
additional factor involving the momentum associated with the
corresponding component. Additional factors involving powers of $p_u$,
$p_v$, $p_y^2$ and $p_z^2$ (odd powers of $p_y$ or $p_z$ give a
vanishing contribution) leave the argument employed in the previous
paragraph unchanged and the same conclusions obtained when $\sigma_u
\to 0$ hold for the smeared noise kernel as well (including the
differential operators). For example we have $\mathcal{N}_{vvvv} \sim
1/\sigma_u \sigma_v^5 \sigma_r^2$ for the smeared version of the
$vvvv$ component of the noise kernel.

On the other hand, when both $\sigma_v \neq 0$ and $\sigma_u \neq 0$
the smeared kernel $\mathcal{N}$ is finite even for $\sigma_r =
0$. This can be seen by taking $\sigma_r = 0$ in
Eq.~(\ref{smearing13}). We then have
\begin{eqnarray}
\mathcal{N} &=&
\frac{1}{(1920 \pi)} \frac{1}{(2 \pi)^{4}} \int  d p_v d p_u d^2 p\,
e^{-p_v^2 \sigma_v^2} e^{-p_u^2 \sigma_u^2} \,
\theta(p_v p_u - \vec{p}^{\, 2}) \nonumber \\
&=& \frac{1}{(1920 \pi)} \frac{1}{(2 \pi)^{4}} \int  d p_v d p_u
\pi p_v p_u e^{-p_v^2 \sigma_v^2} e^{-p_u^2 \sigma_u^2}\,
\theta(p_v p_u) \nonumber \\
&=& \frac{2}{(1920 \pi)} \frac{1}{(2 \pi)^{4}} \int_0^\infty  d p_v
\int_0^\infty d p_u \pi p_v p_u e^{-p_v^2 \sigma_v^2}
e^{-p_u^2 \sigma_u^2}
\sim \frac{1}{\sigma_u^2 \sigma_v^2}
\label{smearing15},
\end{eqnarray}
which is finite for $\sigma_u, \sigma_v \neq 0$. It is also clear that
the same conclusion applies to the smeared version of the noise
kernel. For instance we have $\mathcal{N}_{vvvv} \sim 1/\sigma_u^2
\sigma_v^6$.

\subsection{Smearing along a timelike curve and on a spacelike
hypersurface}
\label{sec:timelike}

In this subsection we consider the smeared kernel $\mathcal{N}$
obtained when working with standard cartesian coordinates in Minkowski
spacetime:
\begin{equation}
\mathcal{N} \equiv
\int dt\, d^3 x\, g(t) f(\vec{x})
\int dt' d^3 x' g(t') f(\vec{x}')
N(t-t',\vec{x}-\vec{x}')
\label{smearing21},
\end{equation}
where we used the vector notation for the three spatial components
$x$, $y$ and $z$, and the Gaussian smearing functions are now
\begin{eqnarray}
g(t) = (2 \pi \sigma_t^2)^{-\frac{1}{2}} \exp(-t^2 / 2 \sigma_t^2),
\label{gaussian4}\\
f(\vec{x})  = (2 \pi \sigma_r^2)^{-\frac{3}{2}} \exp(-\vec{x}^2 / 2 \sigma_r^2).
\label{gaussian5}
\end{eqnarray}
Note that $\sigma_r$ corresponds now to the smearing size of the three
spatial directions.
Introducing the semisum and difference variables $T=(t+t')/2$ and
$\Delta_t=t'-t$, Eq.~(\ref{smearing21}) can be rewritten as
\begin{equation}
\mathcal{N} =
\frac{1}{(2 \pi)^4 \sigma_t^2 \sigma_r^6}
\int dT d^3 X e^{-\frac{T^2}{\sigma_t^2}}
e^{- \frac{\vec{X}^2}{\sigma_r^2}}
\int d \Delta_t d^3 \Delta \, e^{-\frac{\Delta_t^2}{4 \sigma_t^2}}
e^{-\frac{\vec{\Delta}^2}{4 \sigma_r^2}} N(\Delta_t,\vec{\Delta})
\label{smearing22}.
\end{equation}
After integrating over $T$ and $\vec{X}$ we have
\begin{equation}
\mathcal{N} =
\frac{1}{(4 \pi)^{2} \sigma_t \sigma_r^3}
\int d \Delta_t d^3 \Delta \, e^{-\frac{\Delta_t^2}{4 \sigma_t^2}}
e^{-\frac{\vec{\Delta}^2}{4 \sigma_r^2}} N(\Delta_t,\vec{\Delta})
\label{smearing23},
\end{equation}
which can be equivalently rewritten in Fourier space as
\begin{equation}
\mathcal{N} =
\frac{1}{(1920 \pi)} \frac{1}{(2 \pi)^{4}} \int  d p_t d^3 p\,
e^{-p_t^2 \sigma_t^2} e^{-\vec{p}^2 \sigma_r^2}\,
\theta(p_t^2 - \vec{p}^{\, 2})
\label{smearing24}.
\end{equation}
A similar argument to that employed in the previous subsection can be
used to show that in this case $\mathcal{N}$ diverges as $\sigma_t
\rightarrow 0$. The integral in Eq.~(\ref{smearing24}) gets two
identical contributions from the intervals $p_t<0$ and $p_t>0$. We
will also take into account that since the integrand is positive, the
integral is also positive and greater than the same integral
restricted to a smaller domain of integration, so that we have
\begin{eqnarray}
\mathcal{N} &\geq&
\frac{2}{(1920 \pi)(2 \pi)^{4}}
\int_{\sigma_r^{-1}}^\infty d p_t e^{-p_t^2 \sigma_t^2}
\int_0^{\sigma_r^{-1}} 4\pi |\vec{p}|^2 d |\vec{p}| e^{-1}
\nonumber \\
&=& \frac{8 \pi e^{-1}}{3 \sigma_r^3} \frac{1}{(1920 \pi)(2 \pi)^{4}}
\int_{\sigma_r^{-1}}^\infty d p_t e^{-p_t^2 \sigma_t^2}
\sim \frac{1}{\sigma_t \sigma_r^3}
\label{smearing25}.
\end{eqnarray}
The last integral is divergent if one takes $\sigma_t \to 0$ (at least
for $\sigma_r \neq 0$ \footnote{The divergence of $\mathcal{N}$ as
$\sigma_t \rightarrow 0$ can also be proven for $\sigma_r=0$
(\emph{i.e.}, in the absence of smearing along the spatial
directions) by taking $\sigma_r=0$ in Eq.~(\ref{smearing24}) and
replacing $\sigma_r^2$ with an arbitrary but fixed positive value in
Eq.~(\ref{smearing25}).}). Thus, $\mathcal{N}$ diverges unless
$\sigma_t \neq 0$.

Using the previous result, it is easy to discuss whether a smeared
version of the actual noise kernel, given by Eq.~(\ref{noise1}), also
diverges when $\sigma_t \to 0$. Each derivative in Eq.~(\ref{noise1})
gives rise to an additional factor involving the momentum associated
with the corresponding component. Additional factors involving powers
of $p_t^2$, $p_x^2$, $p_y^2$ and $p_z^2$ (odd powers of $p_t$, $p_x$,
$p_y$ or $p_z$ give a vanishing contribution) leave the argument
employed in the previous paragraph unchanged and the same conclusions
obtained when $\sigma_t \to 0$ hold for the smeared noise kernel as
well. For example one has $\mathcal{N}_{tttt} \sim 1/\sigma_t
\sigma_r^7$.  This result is in agreement with that obtained in
Ref.~\cite{ford05}

On the other hand, when $\sigma_t \neq 0$ the smeared noise kernel is
finite even for $\sigma_r = 0$. This can be seen by taking $\sigma_r =
0$ in Eq.~(\ref{smearing24}). We then have
\begin{equation}
\mathcal{N} =
\frac{1}{(1920 \pi)} \frac{1}{(2 \pi)^{4}} \int  d p_t d^3 p\,
e^{-p_t^2 \sigma_t^2} \, \theta(p_t^2 - \vec{p}^{\, 2})
= \frac{1}{(1920 \pi)} \frac{1}{(2 \pi)^{4}} \int  d p_t
\frac{4 \pi}{3} p_t^3 e^{-p_t^2 \sigma_t^2}
\sim \frac{1}{\sigma_t^4}
\label{smearing26},
\end{equation}
which is finite for $\sigma_t \neq 0$. It is also clear that the same
conclusion applies to the smeared version of the noise kernel, with
$\mathcal{N}_{tttt} \sim 1/\sigma_t^8$.

\section{Generalization to curved space, arbitrary Hadamard Gaussian
states and general smearing functions}
\label{appB}

The results obtained for the Minkowski vacuum in flat space can be
generalized to curved space and arbitrary Gaussian Hadamard
states. They also apply to more general smearing functions. This will
be shown in this appendix.

The key ingredient is the fact that the Wightman function for any
Hadamard state has the following form in a sufficiently small normal
neighborhood of an arbitrary spacetime:%
\footnote{For a general spacetime it may not be guaranteed that the
series has a non-vanishing radius of convergence rather than being an
asymptotic series \cite{fulling78}. However, for an analytic spacetime
it can be proven that the radius of convergence is non-zero for
globally Hadamard states \cite{fulling78,garabedian64}. We will
restrict ourselves to analytic spacetimes in this appendix, which is
anyway the case for the spacetimes considered in the rest of the
paper.}
\begin{equation}
G^+(x,x') = \frac{u(x,x')}{\sigma_+(x,x')}
+ v(x,x') \ln \sigma_+(x,x') + w(x,x')
\label{hadamard1},
\end{equation}
where $\sigma_+(x,x')$ is the geodetic interval (one half of the
geodesic distance) for the geodesic connecting the pair of points $x$
and $x'$ with an additional small imaginary component added to the
timelike coordinates (this prescription will be defined more precisely
below); $u$, $v$ and $w$ are smooth functions with $v$ and $w$
expandable as
\begin{eqnarray}
v(x,x') = \sum_{n=0}^\infty v_n (x,x') \sigma^n (x,x') ,
\label{hadamard2a} \\
w(x,x') = \sum_{n=0}^\infty w_n (x,x') \sigma^n (x,x')
\label{hadamard2b} .
\end{eqnarray}
where $u(x,x')$, $v_n(x,x')$ and $w_n(x,x')$ satisfy the Hadamard
recursion relations, which uniquely determine $u(x,x')$ and
$v(x,x')$. On the other hand, $w_0 (x,x')$ is not uniquely determined
and contains the information on the particular state that one is
considering, the remaining $w_n (x,x')$ are also determined once a
particular choice of $w_0 (x,x')$ has been made.  Note that for the
Minkowski vacuum in flat space $v(x,x')$ and $w(x,x')$ vanish and
$u(x,x')$ is simply given by a constant. Other Hadamard states in flat
space have a non-vanishing $w(x,x')$ while $u(x,x')$ and $v(x,x')$
remain unchanged.

The biscalar functions $u(x,x')$, $v_n(x,x')$ and $w_n(x,x')$ can in
turn be expanded in the following way:
\begin{equation}
A(x,x') = \sum_{m=0}^\infty A_{a_1 \cdots a_m} (x)\, \sigma^{a_1} \cdots
\sigma^{a_m}
\label{hadamard3},
\end{equation}
where $\sigma^a \equiv g^{ab} \nabla_b \sigma$.
Furthermore, it will be convenient to employ Riemann normal
coordinates, which can always be introduced in a normal
neighborhood. Given a set of normal coordinates $\{y^\mu\}$, the
geodetic interval can be simply expressed as $\sigma(y,y') = (1/2)
(y^\mu - y^{\prime \mu})(y^\nu - y^{\prime \nu}) \eta_{\mu\nu}$. The
prescription for $\sigma_+$ mentioned above corresponds then to
$\sigma_+ (y,y') = (1/2) [ -(y^0 - y^{\prime 0} - i \varepsilon)^2 +
(\vec{y}- \vec{y}')^2 ]$.

Given a smearing function $s(x)$, one can consider the following
smearing of the product of two Wightman functions $N(x,x') =
\mathrm{Re} [G^+(x,x') G^+(x,x')]$:
\begin{equation}
\mathcal{N} = \int d^4 x\, \sqrt{-g(x)} \, s(x)
\int d^4 x'\, \sqrt{-g(x')} \, s(x') \, N(x,x')
\label{smearing30}.
\end{equation}
Next, one changes from a set of absolute coordinates for each one of
the the two points at which the kernel $N(x,x')$ is evaluated to a set
of absolute coordinates for the first point and a set of relative
coordinates for the location of the second point with respect to the
first one. In particular we will choose Riemann normal coordinates for
the relative location of the second point (in order to study the
divergences in the coincidence limit it is sufficient, by taking small
enough smearing sizes, to consider small enough convex neighborhoods
where both normal coordinates can be defined and the series in
Eq.~(\ref{hadamard1}) is convergent). Eq.~(\ref{smearing31}) then
becomes
\begin{equation}
\mathcal{N} = \int d^4 x\, \sqrt{-g(x)} \, s(x)
\int d^4 y\, \sqrt{-\tilde{g}_x (y)} \,
\tilde{s}_x (y) \, \tilde{N}(x,y)
\label{smearing31}.
\end{equation}
For simplicity, in our discussion below we will consider smearing
functions $\tilde{s}_x (y)$ which are Gaussian and independent of
$x$. However, in Sec.~\ref{sec:general_smearing} we will explain how
our results can be extended to more general smearing functions.  We
also note that we will not analyze the integrals in $x$ since those
should be finite (we are considering globally Hadamard states and
regular smearing functions): only the integrals in $y$ are relevant
for the UV divergences associated with the coincidence limit, which
corresponds to $y \to 0$. There could still be IR divergences arising
from the integrals in $x$, but we will be considering smearing
functions which decay sufficiently fast so that this is not the case.

Taking all the previous considerations into account, it becomes clear
that when calculating the smeared kernel $\mathcal{N}$, the most
divergent terms from $N(x,x') = \mathrm{Re} [G^+(x,x') G^+(x,x')]$
will be of the form $1 / \sigma_+^2(x,x')$, $\ln \sigma_+(x,x') /
\sigma_+(x,x')$ or $1 / \sigma_+(x,x')$. When expressed in normal
coordinates, the contribution to $\mathcal{N}$ due to a term of the
first kind, which will be denoted by $\mathcal{N}_1$ and corresponds
to the product of two $1 / \sigma_+(x,x')$ terms, coincides with the
expression for the Minkowski vacuum in flat space. Hence, one can
directly apply the results obtained in Appendix~A. Furthermore, one
expects that the leading divergence to the smeared function
$\mathcal{N}$ in the various limits of vanishing smearing sizes will
come entirely from that kind of terms and that the other terms will
only give rise to subleading divergences. If this is true, the main
conclusions in Appendix~A will also apply to general Hadamard Gaussian
states in curved spacetime.  Let us, therefore, study more carefully
the contributions to $\mathcal{N}$ from the different kinds of
divergent terms and check that this is indeed the case (other possible
divergent terms in addition to the three kinds mentioned above
correspond to multiplying one of those three by some positive power of
$\sigma^a$, and they will be discussed in
Sec.~\ref{sec:finite_terms}).

In order to analyze the contribution from the other two kinds of
divergent terms, denoted by $\mathcal{N}_2$ and $\mathcal{N}_3$, we
will proceed analogously to Appendix~A and work in Fourier space for
the relative normal coordinates. We start by considering a term of the
form $\ln \sigma_+(x,x') / \sigma_+(x,x')$. Using the same Gaussian
smearing functions as in Appendix~A (as explained above, we do not
include any dependence on the absolute coordinate for the first
point), its contribution to $\mathcal{N}$ when smearing around a null
geodesic is given by
\begin{equation}
\mathcal{N}_2 = \frac{1}{(2 \pi)^{4}} \int  d p_v d p_u d^2 p\,
e^{-p_v^2 \sigma_v^2} e^{-p_u^2 \sigma_u^2} e^{-\vec{p}^2 \sigma_r^2} \,
\frac{1}{2} \left[ L(p) + \left(L(-p) \right)^* \right]
\label{smearing32},
\end{equation}
where $L(p_v, p_u, \vec{p}^{\, 2})$ is the Fourier transform of $\ln
(\sigma_+(y,y') /\lambda^2) / \sigma_+(y,y')$ and we need the second
term inside the square brackets because we are interested in the real
part of the product of two Wightman functions.  An explicit expression
for $L(p_v, p_u, \vec{p}^{\, 2})$ can be determined using the
following two Fourier transforms:
\begin{equation}
\frac{1}{\sigma_+(x,x')}
= \frac{1}{\pi} \int  d^4 p\, e^{i p (y-y')}\, \theta(-p^0) \delta(-p^2)
\label{fourier1},
\end{equation}
\begin{eqnarray}
\ln (\sigma_+(y,y') /\lambda^2) = -2 \int  d^4 p\, e^{i p (y-y')}
\lim_{m \rightarrow 0} && \!\! \left\{ -\frac{1}{\pi} \theta(-p^0)
\frac{d}{d m^2} \delta(-p^2 - m^2) \right. \nonumber \\
&& \left.
+ \left[ \frac{1}{2} \ln (2 m^2 \lambda^2) + \gamma -1 \right] \delta^4 (p)
\right\}
\label{fourier2}.
\end{eqnarray}
where Eq.~(\ref{fourier2}) was derived in Appendix~A.2 of
Ref.~\cite{campos95}. We can then write $L(p_v, p_u, \vec{p}^{\, 2})$
as a convolution of those Fourier transforms:
\begin{eqnarray}
L(p) &=& - 4 (2 \pi)^3 \int d^4 q\, \theta(-p^0+q^0))
\delta(-(p-q)^2) \nonumber \\
&&\times \lim_{m \rightarrow 0}  \left\{ -\frac{1}{\pi} \theta(-q^0)
\frac{d}{d m^2} \delta(-q^2 - m^2)
+ \left[ \frac{1}{2} \ln (2 m^2 \lambda^2) + \gamma -1 \right]
\delta^4 (q) \right\}
\label{fourier3}.
\end{eqnarray}
After a certain amount of calculation, one obtains the following
result for $L(p)$:
\begin{equation}
L(p) = 2 (2 \pi)^3 \left\{ \theta(-p^0) \,
\mathcal{P}\!f \! \left[ \theta(-p^2) \frac{1}{p^2} \right]
- \left[ \ln (2 \lambda^2) + \gamma -1 \right] \theta(-p^0) \delta^4 (-p^2)
\right\}
\label{fourier8},
\end{equation}
where $\mathcal{P}\!f$ denotes Hadmard's finite part prescription (a
generalization of the principal value prescription) whose precise
definition can be found in
Refs.~\cite{campos95,schwartz57,zemanian87}.  We will use this result
in the next subsections to compute in Fourier space the contribution
$\mathcal{N}_2$ to the smeared kernel $\mathcal{N}$ for different
kinds of smearings and analyze under what conditions it is finite.

On the other hand, from Eq.~(\ref{fourier1}) one can immediately see
that the contribution from the term $1 / \sigma_+(x,x')$ is simply
given by
\begin{equation}
\mathcal{N}_3 = \frac{1}{2 \pi} \int  d p_v d p_u d^2 p\,
e^{-p_v^2 \sigma_v^2} e^{-p_u^2 \sigma_u^2} e^{-\vec{p}^2 \sigma_r^2} \,
\delta(p_u p_v - \vec{p}^2)
\label{smearing33}.
\end{equation}

\subsection{Smearing around null geodesics}
\label{sec:null2}

\subsubsection{Contributions from $\mathcal{N}_2$ and $\mathcal{N}_3$}
\label{sec:divergent}

As we found in Appendix~A, when considering a smearing of the noise
kernel for the Minkowski vacuum around a null geodesic, $\mathcal{N}$
diverges as $1 / \sigma_u \sigma_v \sigma_r^2$ in the limit of small
$\sigma_u$ (as long as $\sigma_r \neq 0$, otherwise it diverges as $1
/ \sigma_u^2 \sigma_v^2$). Whereas the contribution $\mathcal{N}_1$
from terms of the form $1 / \sigma_+^2(x,x')$ will exhibit the same
divergent behavior, we will show in this subsection that
$\mathcal{N}_2$ and $\mathcal{N}_3$, the other contributions to
$\mathcal{N}$, are arbitrarily smaller than $\mathcal{N}_1$ for
sufficiently small $\sigma_u$ or $\sigma_v$.

Let us start with $\mathcal{N}_3$. First, one rewrites
Eq.~(\ref{smearing33}) as follows:
\begin{eqnarray}
\mathcal{N}_3 &=& \frac{1}{2 \pi} \int  d p_v d p_u d^2 p\,
e^{-p_v^2 \sigma_v^2} e^{-p_u^2 \sigma_u^2} e^{-\vec{p}^2 \sigma_r^2} \,
\delta(p_u p_v - \vec{p}^2) \nonumber \\
&=& \int_0^\infty  d p_v \int_0^\infty d p_u e^{-p_v^2 \sigma_v^2}
e^{-p_u^2 \sigma_u^2} \int_0^\infty d |\vec{p}| \, e^{-|\vec{p}|^2 \sigma_r^2} \,
\delta(\sqrt{p_u p_v} - |\vec{p}|) \nonumber \\
&=& \int_0^\infty  d p_v \int_0^\infty d p_u e^{-p_v^2 \sigma_v^2}
e^{-p_u^2 \sigma_u^2} e^{- p_u p_v \sigma_r^2} \nonumber \\
&=& \int_0^\infty d \xi e^{-\xi \sigma_r^2}
\int_0^\infty \frac{d p_v}{p_v} e^{-p_v^2 \sigma_v^2}
e^{-\frac{\xi^2}{p_v^2} \sigma_u^2}
\label{smearing34},
\end{eqnarray}
where we introduced the new variable $\xi = p_v p_u$ in the last
equality. Next, using the positivity of the integrand, the fact that
the original integral was invariant under interchange of $p_u$ and
$p_v$, and the fact that value of the exponentials is always equal or
less than one, one can derive the following bound for small $\sigma_u$
and $\sigma_v$:
\begin{equation}
\mathcal{N}_3 \leq - \frac{1}{2 \sigma_r^2} (\ln \sigma_u + \ln \sigma_v) + O(1)
\label{smearing36},
\end{equation}
where the higher-order terms involve positive powers of $\sigma_u$ and
$\sigma_v$ (when $\sigma_v$ is not small, one has an expansion only in
terms of $\sigma_u$ and the $\ln \sigma_v$ term is absent). On the
other hand, for $\sigma_r = 0$ the bound is $\mathcal{N}_3 \leq
(\pi^2/8) (1/\sigma_u \sigma_v)$.

Let us now turn our attention to $\mathcal{N}_2$. Substituting
Eq.~(\ref{fourier8}) into Eq.~(\ref{smearing32}), we get
\begin{equation}
\mathcal{N}_2 = - \int  d p_v d p_u d^2 p\,
e^{-p_v^2 \sigma_v^2} e^{-p_u^2 \sigma_u^2} e^{-\vec{p}^2 \sigma_r^2}
\!\! \left\{ \mathcal{P}\!f \! \left[ \theta(p_u p_v - \vec{p}^2) \,
\frac{1}{p_u p_v - \vec{p}^2} \right]
+ \left[ \ln (2 \lambda^2) + \gamma -1 \right]
\delta^4 (p_u p_v - \vec{p}^2) \right\}
\label{smearing37}.
\end{equation}
The contribution from the second term inside the curly brackets has
the same form as $\mathcal{N}_3$. Hence, we need to concentrate on the
first term. After a lengthy calculation one can derive the following
bound:
\begin{equation}
\mathcal{N}_2 \leq L \frac{1}{\sqrt{2 \sigma_u \sigma_v} \, \sigma_r}
+ O (\ln \sigma_u, \ln \sigma_v)
\label{smearing51},
\end{equation}
where $L$ is a constant of order 1. On the other hand, for $\sigma_r =
0$ the bound is $\mathcal{N}_2 \leq L/\sigma_u \sigma_v$. We can see
that $\mathcal{N}_2$ dominates over $\mathcal{N}_3$ in the limit of
small $\sigma_u$ or $\sigma_v$. Nevertheless, it is still
$\mathcal{N}_1$ that provides the leading contribution in that limit.

\subsubsection{Remaining terms}
\label{sec:finite_terms}

The contribution from the remaining terms in $N(x,x')$ to the smeared
function $\mathcal{N}$, which can be seen from terms of the different
kinds that we have already analyzed multiplying them by some positive
powers of $\sigma$ and $\sigma^a$ can be analyzed as follows. If one
is working with Riemann normal coordinates and the corresponding
Fourier variables, each $\sigma$ factor will give rise to a momentum
d'Alembertian $(1/2) \eta_{\mu\nu} (\partial/\partial p_{\mu})
(\partial/\partial p_{\nu})$ acting on the Fourier transform of that
term in $N(x,x')$. Similarly, each $\sigma^\mu$ factor will give rise
to a linear differential operator $(\partial/\partial p_\mu)$ acting
on the Fourier transform. Integrating by parts in the Fourier space
expression for $\mathcal{N}$ so that the d'Alembertian acts on the
Fourier transform of the smearing functions, it will produce a factor
of the form $(-\sigma_u^2 \sigma_v^2 p_u p_v + 2 \sigma_r^4 \vec{p}^2
-2 \sigma_r^2)$. Proceeding analogously with the linear operator
$(\partial/\partial p_\mu)$, one gets the factors $2 \sigma_u^2 \,
p_u$, $2 \sigma_v^2 \, p_v$ and $2 \sigma_r^2 \, p_j$ (where the index
$j$ corresponds to one of the two orthogonal spatial components) when
$\mu$ equals $u$, $v$ or $j$ respectively.

Next, one needs to see how the results of the integrals in
Sec.~\ref{sec:divergent} change when the integrand is multiplied by
positive powers of $p_u$, $p_v$ and $p_j$. Odd powers of $p_j$ for any
$j$ give a vanishing result since the integrals and integrands there
are symmetric under a sign change of $p_j$, whereas every $(p_j)^2$
factor can be written as $(-1/2) (\partial / \partial
\sigma_r^2)$. Similarly, every factor $p_u^2$ or $p_v^2$ can be
written as $(-1/2) (\partial / \partial \sigma_u^2)$ or $(-1/2)
(\partial / \partial \sigma_v^2)$ respectively.%
\footnote{In general when two functions satisfy a certain inequality
  that does not imply that their derivatives will satisfy it. This
  will, however, be the case when considering the bounds derived in
  the previous sections in order to analyze the leading divergent
  behavior in the limit $\sigma_u \to 0$.}  On the other hand, having
an odd power of $p_u$ only or $p_v$ only gives a vanishing result
because the integrals and integrands in Sec.~\ref{sec:divergent} are
symmetric under a simultaneous sign change of $p_u$ and
$p_v$. However, odd powers of $p_u p_v$ do not vanish in general
because the integrands are not symmetric under interchange of $p_u$ or
$p_v$ only. Therefore, one needs to check how the main results for the
integrals in Sec.~\ref{sec:divergent} change when the integrands are
multiplied by a $p_u p_v$ factor. In Eq.~(\ref{smearing34}) it gives
rise to a factor $\xi$, which implies an additional $1 / \sigma_r^2$
factor multiplying the final result for $\mathcal{N}_3$ in
Eq.~(\ref{smearing36}). One can similarly find that an additional $p_u
p_v$ factor in Eq.~(\ref{smearing37}) also implies a $1 / \sigma_r^2$
factor multiplying the final result for $\mathcal{N}_2$ in
Eq.~(\ref{smearing51}).

We are finally in a position to discuss the effects of $\sigma$ and
$\sigma_a$ factors multiplying the contributions to the smeared kernel
$\mathcal{N}$. Each power of $p_u^2$ and $p_v^2$ (including their
$\sigma_u^4$ and $\sigma_v^4$ accompanying factors) will typically
give rise to $\sigma_u^2$ and $\sigma_v^2$ factors respectively. A
$p_u p_v$ factor (with its accompanying $\sigma_u^2 \sigma_v^2$
factor) will give rise to a $\sigma_u^2 \sigma_v^2 / \sigma_r^2$
factor. And each $p_j^2$ factor (with its accompanying $\sigma_r^4$
factor) will give rise to a $\sigma_r^2$ factor. Thus, we see that the
divergent behavior in the limit of small $\sigma_u$ remains unchanged
or even gets improved. (An analogous conclusion would apply in the
limit of small $\sigma_r$.) It then follows that the behavior of
$\mathcal{N}$ in the small $\sigma_u$ limit is still dominated by the
flat space vacuum contribution $\mathcal{N}_1$.

\subsubsection{Smearing of the actual noise kernel}
\label{sec:actual_noise}

The actual noise kernel involves a number of functions multiplying the
kernel $N(x,x') = \mathrm{Re} [G^+(x,x') G^+(x,x')]$ and differential
operators acting on it. When using relative Riemann normal coordinates
for the location of the second point with respect to the first one,
the part of these linear operators that depends on the relative
coordinates of the second point can be entirely expressed in terms of
functions and tensor fields (such as the metric and the curvature
tensors) as well as partial derivatives. That is not the case in
general for the operators associated with the first spacetime point,
but the dependence on the first point does not exhibit a divergent UV
behavior.

The functions multiplying $N(x,x')$ can be expanded in terms of the
relative Riemann normal coordinates, which involves powers of $\sigma$
and $\sigma^a$ that can be treated as explained in the previous
subsubsection. As we saw, they either leave the divergent behavior in
the limit of small $\sigma_u$ unchanged or decrease the degree of
divergence.

On the other hand, the partial derivatives $(\partial/\partial y^\mu)$
simply correspond to $i p_\mu$ factors in Fourier space, which can
also be dealt with as discussed in the previous subsubsection. Even
powers of $p_u$ increase the degree of divergence in the limit of
small $\sigma_u$: every $p_u^2$ factor gives rise to a $1 /
\sigma_u^2$ factor. All others leave the degree of divergence
unchanged. Moreover, since the $p_u^2$ affect both the kind of terms
contributing to $\mathcal{N}_1$ and those contributing to
$\mathcal{N}_2$ and $\mathcal{N}_3$, the conclusion that the leading
divergent behavior when $\sigma_u \to 0$ is given by the Minkowski
vacuum result remains unchanged for the actual noise kernel.

\subsection{Smearing along timelike geodesics and on spacelike
hypersurfaces}
\label{sec:timelike2}

The results for smearings of the noise kernel on spacelike
hypersurfaces and along timelike geodesics obtained for the vacuum
state in Minkowski can be generalized proceeding analogously to what
was done in the previous subsection for smearings along null
geodesics.

Let us start by considering $\mathcal{N}_3$:
\begin{eqnarray}
\mathcal{N}_3 &=& \frac{1}{2 \pi} \int  d p_t d^3 p\,
e^{-p_t^2 \sigma_t^2} e^{-\vec{p}^2 \sigma_r^2} \,
\delta(p_t^2 - \vec{p}^2) \nonumber \\
&=& 4 \int_0^\infty  d p_t  e^{-p_t^2 \sigma_t^2}
\int_0^\infty d |\vec{p}| \, |\vec{p}| e^{-|\vec{p}|^2 \sigma_r^2} \,
\delta(p_t - |\vec{p}|) \nonumber \\
&=& 4 \int_0^\infty  d p_t  \, p_t e^{-p_t^2 \sigma_t^2}
e^{-p_t^2 \sigma_r^2}
\label{smearing53}.
\end{eqnarray}
Hence, we have
\begin{equation}
\mathcal{N}_3  = \frac{2}{\sigma_t^2 + \sigma_r^2}
\label{smearing54},
\end{equation}
which is finite provided that $\sigma_t \neq 0$ or $\sigma_r \neq 0$.

Let us now turn our attention to $\mathcal{N}_2$. Applying the result
in Eq.~(\ref{fourier8}) to this case, we get
\begin{equation}
\mathcal{N}_2 = - \int  d p_t d^3 p\,
e^{-p_t^2 \sigma_t^2} e^{-\vec{p}^2 \sigma_r^2}
\!\!  \left\{ \mathcal{P}\!f \! \left[ \theta(p_t^2 - \vec{p}^2) \,
\frac{1}{p_t^2 - \vec{p}^2} \right]
+ \left[ \ln (2 \lambda^2) + \gamma -1 \right]
\delta^4 (p_t^2 - \vec{p}^2) \right\}
\label{smearing55}.
\end{equation}
The contribution from the second term inside the curly brackets has
the same form as $\mathcal{N}_3$. Hence, we only need to concentrate
on the first term. After a slightly lengthy calculation, one gets the
following bound for $\mathcal{N}_2$ (when $\sigma_r \neq 0$):
\begin{equation}
\mathcal{N}_2 < \frac{C_1}{\sigma_r^2} + \frac{C_2}{\sigma_t^2}
\label{smearing64},
\end{equation}
where $C_1$ and $C_2$ are positive dimensionless constants which are
finite provided that $\sigma_t \neq 0$ (they behave like $- \ln
\sigma_t$ in the limit of small $\sigma_t$). On the other hand, for
$\sigma_r = 0$ one has a bound given by
\begin{equation}
|\mathcal{N}_2| < \frac{D}{\sigma_t^2}
\label{smearing66},
\end{equation}
where $D$ is some positive dimensionless constant which is finite
provided that $\sigma_t \neq 0$ (it behaves like $- \ln \sigma_t$ in
the limit of small $\sigma_t$). This shows that just a temporal
smearing is enough to render $\mathcal{N}_2$ finite. Of course
$\mathcal{N}_2$ will also be finite if, in addition to the temporal
smearing, some but not all of the (orthogonal) spatial directions have
a non-vanishing smearing size.

Proceeding similarly to
Secs.~\ref{sec:finite_terms}-\ref{sec:actual_noise}, one can argue
that factors involving positive powers of $\sigma$ and $\sigma^a$ as
well as the derivative operators in configuration space do not alter
the main conclusion in this subsection. Thus, a smearing along the
timelike direction is enough to have a finite result for the smeared
versions of $N(x,x')$ and the actual noise kernel.

\subsection{More general smearing functions}
\label{sec:general_smearing}

In Secs.~\ref{sec:null2} and \ref{sec:timelike2} we considered
Gaussian smearing functions for the relative normal
coordinates. However, as pointed out at the beginning of this
appendix, when transforming from a pair of sets of \emph{absolute}
coordinates for the two points where the noise kernel is evaluated to
one set of \emph{absolute} coordinates and one set of \emph{relative}
ones, the form of the smearing functions will change in general. That
will also happen when transforming to Riemann normal coordinates if
one had initially chosen a different kind of relative
coordinates. Furthermore, even in the flat space case one may be
interested in considering other kinds of smearing functions. For
instance, one may wish to consider a smearing function adapted to a
spherical surface rather than a plane.

The results obtained in these appendices can be extended to more
general smearing functions, making it possible to cover the situations
described in the previous paragraph. The essential idea is simple: the
previous results can be generalized for smearing functions which can
be locally approximated by the Gaussian smearing functions in Riemann
normal coordinates considered so far. More specifically, by ``locally
approximated'' we mean that in those coordinates the new smearing
functions can be expressed as the Gaussian smearing functions times a
factor involving an expansion in positive powers of $\sigma$ and
$\sigma^a$. The procedure described in Sec.~\ref{sec:finite_terms} can
then be employed to show that the main results remain
unchanged. Moreover, the detailed form of the smearing functions for
large values of the Riemann normal coordinates (or even outside the
range where they can be defined) is not relevant when concerned with
the divergent behavior (when certain smearing sizes tend to zero) as
we are are here. Or put in a different way, even though in general the
form of the smearing function can be significantly distorted when
transforming from the original coordinates to relative Riemann normal
coordinates, this effect becomes less and less important when
considering sufficiently small smearing sizes (which is in any case
the relevant regime to study the UV divergent behavior): for
sufficiently small scales the coordinate transformation is
characterized by the linear map between tangent spaces (the Jacobian
matrix evaluated in the coincidence limit) plus small corrections
involving positive powers of $\sigma$ and $\sigma^a$, which can be
dealt with following the procedure described in
Sec.~\ref{sec:finite_terms}.

These points can be illustrated with a simple example. Consider a
spatial smearing function adapted to a sphere in flat space with the
following form (in spherical coordinates):
\begin{equation}
f(r,\theta)  = \frac{1}{(2 \pi)^{3/2} \sigma_r r_0^2 \sigma_\theta^2 \
 b(\sigma_\theta)}
\exp \left[-\frac{(r-r_0)^2}{2 \sigma_r^2} \right]
\exp \left[-\frac{\theta^2}{2 \sigma_\theta^2} \right]
\label{gaussian6},
\end{equation}
where $b(\sigma_\theta)$ is some dimensionless function which ensures
the proper normalization of the smearing function and tends to $1$
when $\sigma_\theta \to 0$ [note also that for $\theta$ we actually
have half a Gaussian since its domain is $(0,\pi)$].  This can be
written in terms of Riemann normal coordinates (cartesian coordinates
in this case) adapted to the plane tangent to the sphere at the point
$(r=r_0,\ \theta=0)$ as follows:
\begin{eqnarray}
f(x,y,z)  &=& \frac{1}{(2 \pi)^{3/2} \sigma_r r_0^2 \sigma_\theta^2 \,
b(\sigma_\theta)}
\exp \left[-\frac{(z-r_0)^2}{2 \sigma_r^2} \right]
\exp \left[-\frac{x^2 + y^2}{2 r_0^2 \sigma_\theta^2} \right]
\nonumber \\
&&\times \left[ 1 + O\left( \frac{(z-r_0)^3}{r_0^3 \sigma_\theta^2},
\frac{(x^2 + y^2)(z-r_0)}{r_0^3 \sigma_\theta^2},
\frac{(x^2 + y^2)(z-r_0)}{r_0 \sigma_r^2}
\right)  \right]
\label{gaussian7}.
\end{eqnarray}
We can see that for any given $r_0 > 0$, if one chooses sufficiently
small $\sigma_r$ and $\sigma_\theta$, the higher-order terms become
negligible (in the region where there is not a large suppression due
to the exponential factors they are very small). Thus, $f(x,y,z)$
corresponds to a Gaussian smearing function in cartesian coordinates
with $\sigma_x = \sigma_y = r_0 \sigma_\theta$ and $\sigma_z =
\sigma_r$ plus small corrections.


\end{document}